\documentclass[aps,prb,twocolumn,showpacs,preprintnumbers,amsmath,amssymb,superscriptaddress]{revtex4-1}
\usepackage{dcolumn}
\usepackage{bm}
\usepackage{graphicx}
\usepackage{times}
\usepackage{epstopdf}
\usepackage{color}
\usepackage[subrefformat=parens,labelformat=parens,caption=false]{subfig}
\usepackage[pdftex,colorlinks,bookmarks = true]{hyperref} 
\usepackage[capitalize]{cleveref} 

\begin{document}

\definecolor{Red}{rgb}{1,0,0}
\definecolor{Blu}{rgb}{0,0,01}
\definecolor{Green}{rgb}{0,1,0}
\definecolor{Purple}{rgb}{0.5,0,0.5}
\newcommand{\red}{\color{Red}}
\newcommand{\blu}{\color{Blu}}
\newcommand{\green}{\color{Green}}
\newcommand{\purple}{\color{Purple}}
\newcommand{\rvec}{\mathbf{r}}
\newcommand{\nvec}{\mathbf{n}}
\newcommand{\ev}[1]{\langle#1\rangle}
\newcommand{\abs}[1]{\left|#1\right|}
\newcommand{\uone}{\ensuremath{\mathrm{U}(1)} }
\newcommand{\sutwo}{\ensuremath{\mathrm{SU}(2)} }
\newcommand{\ztwo}{\ensuremath{\mathbb{Z}_2} }
\newcommand{\im}{\mathrm{i}}
\newcommand{\vecn}{\ensuremath{\mathbf{n}} }
\newcommand{\bv}[1]{\boldsymbol{#1}}
\newcommand{\bvh}[1]{\hat{\bv{#1}}}
\newcommand{\dd}[0]{\text{d}}

\title{Competing interactions in population-imbalanced two-component Bose-Einstein condensates}
\author{Peder Notto Galteland}
\affiliation{Department of Physics, NTNU, Norwegian University of Science and Technology, N-7491
Trondheim, Norway}
\author{Asle Sudb\o}
\affiliation{Department of Physics, NTNU, Norwegian University of Science and Technology, N-7491
Trondheim, Norway}
\date{\today}

\begin{abstract}
We consider a two-component Bose-Einstein condensate with and without synthetic "spin-orbit"
interactions in two dimensions. Density- and phase-fluctuations of the condensate are included,
allowing us to study the impact  of thermal fluctuations and density-density interactions on the
physics originating with spin-orbit interactions. In the absence of spin-orbit interactions, we find
that inter-component density interactions deplete the minority condensate. The thermally driven
phase transition is driven by coupled density and phase-fluctuations, but is nevertheless shown to
be a phase-transition in the Kosterlitz-Thouless universality class with close to universal
amplitude ratios irrespective of whether both the minority- and majority condensates exist in the
ground state, or only one condensate exists.
In the presence of spin-orbit interactions we observe three separate phases, depending on the
strength of the spin-orbit coupling and inter-component density-density interactions: a
phase-modulated phase with uniform amplitudes for small intercomponent interactions,
a completely imbalanced, effectively single-component, condensate for intermediate spin-orbit
coupling strength and suficciently large inter-component interactions, and a phase-modulated
\textit{and} amplitude-modulated phase for sufficiently large values of both the spin-orbit coupling
and the inter-component density-density interactions.
The phase which is modulated by a single $\bv q$-vector only is observed to transition into an isoptropic
liquid through a strong de-pinning transition with periodic boundary conditions, which weakens with
open boundaries.
\end{abstract}
\maketitle

\section{Introduction}
\label{sec:intro}

Spin-orbit coupling (SOC) underpins many fascinating phenomena in condensed matter physics,
including the spin-Hall\cite{Kato2004,Konig2007} effects and the existence of topological
insulators\cite{Kane2005,Bernevig2006,Hsieh2008,Hasan2010}. SOC is also important for determining
the physical properties of such important functional materials as GaAS\cite{Koralek2009}. Due to the
fundamental magneto-electric character of SOC in charged systems, it also has important ramification
for the manipulation of spin-degrees of freedom using electric fields, currently a research topic of
intense focus. While these examples represent systems  where a real physical spin is coupled to the
orbital motion of electrons, similar phenomena may also be investigated in bosonic systems. Here,
the SOC does not originate with a relativistic correction to the equations of motion, as they do in
the electronic systems mentioned above. Rather, they are synthetic in the sense of being
engineered\cite{Lin2011,Galitski2013} Rashba\cite{Bychkov1984}- and
Dresselhaus\cite{Dresselhaus1955} couplings in multi-component Bose-Einstein condensates. Such
multi-component condensates could be either homo-nuclear with different species occupying different
hyperfine spin states\cite{Myatt1997,Hall1998}, or they could be mixtures of different types of
bosons\cite{Modugno2002,McCarron2011}. In either case, one may associate an index with each species
of the condensate, serving as an internal "`spin"-degree of freedom. A great advantage of studying
the physics of competing interactions and couplings in Bose-Einstein condensates or other ultra-cold
atomic systems is that the interaction parameters, namely density-density interactions and
"spin-orbit" couplings, are highly tunable. This facilitates the study of a wide range of phenomena
otherwise not accessible in standard condensed matter systems.

SOC in a confined bosonic gas of cold atoms has been achieved using an optical Raman-dressing
scheme\cite{Lin2011}. A similar scheme has also been used in cold fermionic gases\cite{Wang2012}.
In optical lattices\cite{Bloch2005} a synthetic SOC has been realized in a
one-dimensional lattice using a similar Raman-dressing scheme\cite{Hamner2015}. Other proposals for
realizing SOC in an optical lattice include periodically driving the lattice with an oscillating magnetic
field gradient\cite{Struck2014}, or by using off-resonance laser beams\cite{Kennedy2013}. The two
latter schemes avoids the problem of heating caused by spontaneous emission of photons as they do
not rely on near-resonant laser fields.

In the case of topological insulators, the classification scheme and the physical properties of
these systems are largely worked out and predicted at zero temperature and ignoring many-body
interactions\cite{Schnyder2008,Kitaev2009,Qi2009,Hasan2010}. It seems worthwhile to examine the
effects of both temperature and many-body interactions on the effect of SOC. In this respect,
looking at "pseudo-spin" Bose-Einstein condensates offers an attractive alternative for studying
many-body effects, since one can, among other things, perform large-scale Monte-Carlo simulations
without the complicating factors arising from Fermi-statistics in the problem. Bosonic systems also
have the attractive property of featuring a condensate at low enough temperatures, such that one has
a mean-field starting point to compare with, at least provided the system is placed far enough away
in parameter space from critical point arising either from interaction effects or thermal
fluctuations.

Previous works on bosonic spin-orbit coupled condensates have shown that their ground state has a
periodically modulated striped spin structure both in a lattice
model\cite{Gras2011,Cole2012,Radic2012,Toniolo2014} and by considering the continuum Gross-Pitaevskii
equations\cite{Stanescu2008,Yip2011,Kasamatsu2015}. Including SOC splits the energy bands of spin-up and
spin-down particles into bands of definite helicity, where the lower band will have minima at finite
momentum, provided that any additional Zeeman-splitting (i.e imbalance in the condensate density) of
the bands is not too large. A continuum model will have a degenerate ring of minima in momentum
space, with fixed length of the momentum-vector in two dimensions, while a square lattice will break
the degeneracy down to four points along the diagonals of the lattice. It has also been shown that,
in the weak coupling limit, the bosons will condense either into one or two minima in the ground
state, depending on the strength of the intra-component interactions. {Furthermore, the
stability of the of Rashba-coupled Bose gases in the presence of thermal and quantum fluctuations
has been studied in the Bogoliubov approximation\cite{Barnett2012,Ozawa2012}} However, the full
range of thermal fluctuations has not been considered before in such systems.

In this paper, we therefore consider a two-dimensional two-component Bose-Einstein condensate with a
Rashba synthetic SOC. The condensates are also assumed to be population-imbalanced with different
densities among the components in the ground state. Fluctuation effects are strong in
two-dimensional system such that no local order parameters exist for systems with continuous
symmetries. Even so, one may get some rough insights into the effects of varying interactions and
temperature at the mean-field level. For a spin-orbit coupled system featuring a non-uniform ground
state, this differs from the case where one expects a uniform ground state, in that the gradient
terms of the theory need to be included even at the mean-field level. We will perform such a
mean-field analysis in this paper, and compare the results to what we obtain in large-scale
Monte-Carlo simulations.  At low temperatures, we find that that a mean-field analysis yields
results for critical values of interaction parameters that destroy the minority-condensate in good
agreement with Monte-Carlo simulations. At elevated temperatures, we find that the
amplitude-fluctuating two-component condensate undergoes Kosterlitz-Thouless phase transitions for
two qualitatively different parameter regimes. \textit{i)} In the absence of SOC, we find that the
condensate loses phase-coherence via proliferation of vortex-antivortex pairs in an
amplitude-fluctuating background, and that this phase transition is a Kosterlitz-Thouless phase
transition with a universal amplitude ratio of the jump in superfluid density to critical
temperature given by $2/\pi$. \textit{ii)} In a parameter regime where SOC plays a role, and gives a
non-uniform ground state in the form of stripes of modulated {\it phases} (but not amplitudes) of
the condensate ordering fields, we find via finite-size scaling of the structure functions at the
pseudo-Bragg vectors that the stripes melt through thermal depinning from the lattice, and not in a
Kosterlitz-Thouless phase transition.  When the condensate we study features a
non-uniform ground state, it may be thought of as a bosonic analogue to either a two-dimensional
two-component superconductor in a Larkin-Ovchinnikov state, or to a one-component superconductor in
a Fulde-Ferrell state. The former features topological order at finite temperature, the latter is
topologically disordered at any finite temperature \cite{Radzihovsky2011}.

The paper is structured as follows. \cref{sec:model} presents the model and observables we use to
classify the states and transitions observed. \cref{sec:mft} contains the mean-field calculations.
\cref{sec:MCdetails} describes the Monte-Carlo scheme we use. In \cref{sec:results} all our
Monte-Carlo results as well as discussions of their significance is included. We present our
conclusions in \cref{sec:conclusions}.

\section{Model}
\label{sec:model}

In this section, we present the lattice model used in the Monte-Carlo simulations, discuss some of
its basic properties and present the observables measured in simulations to classify the phases and
phase transitions we observe.

\subsection{Ginzburg-Landau model}
The starting point of our formulation is the standard two-component Ginzburg-Landau model with an
added SOC, given by
\begin{align}
  H =  \int\textit{d}^2r\Bigg[&\frac{1}{2}\abs{\nabla\Psi}^2 + V(\Psi)\Bigg] +H_\text{SO}
\label{eq:GL_modelbase}
\end{align}
Here, $\Psi^\dag=(\psi_1^*, \psi_2^*)$ is a spinor of two complex fields, where the individual
components may be though of as a pseudospin degree of freedom, and $V$ is the potential. We
allow the potential to contain inter- and intra-component density-density interactions, as
well as a chemical potential. The chemical potential is chosen to have different strengths
for each component, which may be viewed as a Zeeman-like field acting on the pseudospins.
\begin{equation}
  V(\Psi) = \sum_i\alpha_i |\psi_i|^2 + \sum_{ij}g_{ij} |\psi_i|^2 |\psi_j|^2
\end{equation}
The term containing the spin-orbit interaction, $H_\text{SO}$, is of the Rashba type,
on the form
\begin{equation}
  H_\text{SO} =
  \frac{i\kappa}{2}\int\textit{d}^2r\Psi^\dag\big((\bv\sigma\times\nabla)\cdot\bvh
z\Big)\Psi+\text{h.c.}
\end{equation}
We may write the SOC on component form
\begin{align}
  H_\text{SO} =
  \frac{\kappa}{2}\int\textit{d}^2r\Bigg[&\psi_2^*\partial_x\psi_1-\psi_1^*\partial_x\psi_2\nonumber\\
                                          &+i\psi_2^*\partial_y\psi_1+i\psi_1^*\partial_y\psi_2\Bigg]+\text{h.c.}
  \label{eq:Hsoc_comp}
\end{align}
To simplify the representation of the potential term, we introduce the following parametrization
$\alpha_1 = \alpha(1-\Delta),\;\alpha_2 = \alpha(1+\Delta),\;g_{11} = g(1-\gamma),\;g_{22} =
g(1+\gamma),\;g_{12} = \lambda g$. $\Delta$ thus tunes the imbalance of the components, $\gamma$
tunes the relative strengths of the intra-component density-density interactions, while $\lambda$
tunes the strength of the inter-component density-density interaction. The latter is responsible
for producing a phase-separated state.

{
This particular Ginzburg-Landau theory has been much studied in the literature in the absence 
of SOC. It features a rich phase diagram where either one or both of the condensates may exist. 
In three dimensions, the phases are separated by first- or second-order phase transitions, 
depending on the details of the model\cite{Ceccarelli2016}. The main impact of SOC is to 
produce a qualitatively new feature compared to the case without SOC, namely a non-uniform 
ground state, see below. 
}

\subsection{Lattice formulation}

To arrive at a lattice model suitable for Monte-Carlo simulations, we discretize the continuous
fields $\psi_i$ on a square grid, that is we let $\psi_i\rightarrow\psi_{\bv r,
i}$, where $\bv r = (r_x, r_y)$, $r_\mu\in (1, \ldots, L)$ and $\mu \in(x, y)$. The
derivatives are converted to forward finite differences through the replacement
\begin{equation}
  \partial_\mu\psi_i\rightarrow \frac{1}{a}\left(\psi_{\bv r+\bvh\mu, i}-\psi_{\bv r, i}\right),
  \label{eq:forwdiff}
\end{equation}
where $\bvh\mu$ is a unit vector in the $\mu$-direction, and $a$ is the lattice spacing.  We
suppress the lattice spacing in the following expressions, real space distances are plotted in units
of $a$, while reciprocal space is plotted in units of $2\pi/La$. By introducing real amplitudes and
phases, $\psi_{\bv r, i} = \left|\psi_{\bv r, i}\right|\exp(i\theta_{\bv r, i})$ we may write the
derivatives of the Hamiltonian in terms of trigonometric functions.

We write the Hamiltonian as a sum of three terms as follows
\begin{equation}
  H  = H_K + H_{\text{SO}}  + H_V.
  \label{eq:Hlattice}
\end{equation}
$H_K$ contains the kinetic terms, which are written in the standard cosine formulation
\begin{align}
  H_K =
  \sum_{\rvec,\bvh\mu,i}\left(\abs{\psi_{\rvec}}^2-\abs{\psi_{\rvec+\bvh\mu,i}}\abs{\psi_{\rvec,i}}\cos\Delta_\mu\theta_{\rvec,i}\right).
  \label{eq:Hklattice}
\end{align}
The potential term, with the new parametrization, is now written as
\begin{align}
  H_V = \sum_{\rvec}\Bigg[&-\alpha(1-\Delta)\left|\psi_{\bv r,
    1}\right|^2-\alpha(1+\Delta)\left|\psi_{\bv r, 2}\right|^2\nonumber\\
    &+g(1-\gamma)\left|\psi_{\bv r, 1}\right|^4+g(1+\gamma)\left|\psi_{\bv r, 2}\right|^4\nonumber\\
    &+2g\lambda\left|\psi_{\bv r, 1}\right|^2\left|\psi_{\bv r, 2}\right|^2\Bigg]
  \label{eq:HVlattice}
\end{align}
The SOC term on the lattice may also be described in terms of trigonometric functions. By
replacing the differential operators of \cref{eq:Hsoc_comp} by the forward difference
representation of
\cref{eq:forwdiff}, and then replacing the complex fields with the amplitude and phase
representation, we may write this particular term of the Hamiltonian as
\begin{align}
  H_{\text{SO}} = -\kappa\sum_\rvec\Bigg[&\abs{\psi_{\bv r, 1}}\abs{\psi_{\bv r + \bvh x,
  2}}\cos(\theta_{\bv r +\bvh x, 2}-\theta_{\bv r, 1})\nonumber\\
-&\abs{\psi_{\bv r, 2}}\abs{\psi_{\bv r+\bvh x, 1}}\cos(\theta_{\bv r + \bvh x, 1}-\theta_{\bv r, 2})\nonumber\\
+&\abs{\psi_{\bv r, 1}}\abs{\psi_{\bv r + \bvh y, 2}}\sin(\theta_{\bv r + \bvh y, 2}-\theta_{\bv r, 1})\nonumber\\
+&\abs{\psi_{\bv r, 2}}\abs{\psi_{\bv r + \bvh y, 1}}\sin(\theta_{\bv r + \bvh y, 1}-\theta_{\bv r, 2}) \Bigg].
  \label{eq:HSOlattice}
\end{align}

\subsection{London model}

Thermal fluctuations of the phases of the complex order parameter component
are the most relevant fluctuations. Hence, it is useful first to neglect the amplitude
fluctuations and consider a London-model of the problem. To this end, we write the complex fields as
$\psi_i=\rho_i\exp{i\theta_i}$, where only the phase $\theta_i$ is allowed to fluctuate. Note that
this also implies that we assume the amplitudes to be uniform. To arrive at a London formulation
we write the Ginzburg-Landau Hamiltonian of \cref{eq:GL_modelbase} on component form, and
replace the complex fields with a constant amplitude and a fluctuating phase, as described above.
This gives
\begin{align}
  H = \int\textit{d}^2r\Bigg[\sum_i\frac{\rho_i^2}{2}&(\nabla\theta_i)^2\nonumber\\
  -\kappa\rho_1\rho_2\Big[&\sin(\theta_1-\theta_2)\partial_x(\theta_1+\theta_2)\nonumber\\
  +&\cos(\theta_1-\theta_2)\partial_y(\theta_1+\theta_2)\Big]\Bigg],
  \label{eq:HSOLondon}
\end{align}
such that two composite variables with very different behaviors emerge. On the one hand,
$\theta_-\equiv\theta_1-\theta_2$ has a preferential value: in the presence of the gradients of the
phase  sum the second term in the above equations has phase-locking effects.  On the other hand,
$\theta_+\equiv\theta_1+\theta_2$ has first order gradient terms, which may make it energetically
favorable to modulate this phase. As the SOC-term couples the two variables, there may be subtle
interplay between them influencing the phase transitions of the model.

{
The scaling dimension of the SOC-term will be one less than that of a Josephson coupling 
(a Josephson term has no derivatives, while the SOS-term has a single derivative). The 
SOC coupling is therefore less relevant, in renormalization-group sense, than the 
Josephson-coupling. A Josephson-coupling is a singular perturbation on the system where 
Josephson-coupling is absent, being highly relevant at any strength of the coupling
(see for instance Appendix E of Ref. \onlinecite{PhysRevB.71.214509}, in particular the 
discussion following Eq. E7).  It 
leads to a locking of phases of the complex order-parameters of each component of the 
condensate, thus reducing the symmetry of the system from $U(1) \times U(1)$ to $U(1)$ 
(for the two-component case we study in this paper). On the other hand, the scaling 
dimension of the SOC term is one higher than current-current interactions in a 
multi-component BEC, a so-called Andreev-Bashkin term 
\cite{PhysRevA.72.013616,PhysRevB.78.144510}, 
which leaves the $U(1) \times U(1)$ symmetry of the uncoupled system intact. The SOC 
coupling is an interesting case falling in between these two cases. Namely, at given 
imbalance, a critical value of the SOC must 
be reached before the SOC term leads to a non-uniform ground state. Below this critical
strength the system effectively is represented (ignoring for the moment many-body 
interactions) as two independent condensates with $U(1) \times U(1)$ symmetry. Above
the critical value of SOC, the system takes up a finite-momentum ground state. The
SOC then effectively acts as a finite-momentum phase-locking Josephson coupling, 
as we shall see below. 
}

Below, we will perform a mean-field analysis, where we assume that the phases and amplitudes of the
boson condensate are modulated by some wave vector, which is included as a variational parameter
when the free energy is minimized. This result may be compared to the previous work done on SOC
bosons. We also compare the mean-field analysis to Monte-Carlo simulations of the interacting
lattice model.

\subsection{Observables}

The phase transition observed at $\kappa=0$ is classified by examining the helicity modulus,
defined by
\begin{equation}
  \ev{\Upsilon_{i, \mu}} \equiv \frac{1}{V}\frac{\partial^2 F(\Delta_{i, \mu})}{\partial\Delta_{i, \mu}^2},
\end{equation}
along with the fourth order modulus
\begin{equation}
  \ev{\Upsilon_{4, i, \mu}} \equiv \frac{1}{V^2}\frac{\partial^4 F(\Delta_{i,
  \mu})}{\partial\Delta_{i, \mu}^4}.
\end{equation}
where $\Delta_{i, \mu}$ is an infinitesimal twist applied to the phase $\theta_{r, i}$ in the
$\mu$-direction and $F(\Delta_{i, \mu})$ is the free energy with this twist applied. The transition
manifests itself as a discontinuity in the helicity modulus in the thermodynamic limit. This
translates to a dip in the fourth order modulus that does not vanish in the thermodynamic limit. See
\cref{app:KTclass} for more details. As the $x$- and $y$-directions are equivalent, we will consider
the average of $\Upsilon_{i, x}$ and $\Upsilon_{i, y}$ denoted by $\Upsilon_{i, \perp}$, as well as
the average of $\Upsilon_{4, i, x}$ and $\Upsilon_{4, i, y}$ denoted by $\Upsilon_{4, i, \perp}$.

To examine the thermal melting of the spin-orbit induced ground-state modulation, we calculate the
specific heat, $C_v$. It is given as fluctuations of the Hamiltonian
\begin{equation}
  C_V = \beta^2\left(\left\langle H^2\right\rangle-\left\langle H\right\rangle^2\right)
\end{equation}

To compare the Monte-Carlo results to mean-field calculations, we measure the average amplitude
$u_i$, defined as
\begin{equation}
  u_i = \Bigg\langle\sum_{\bv r}\abs{\psi_{i, r}}^2\Bigg\rangle.
\end{equation}
Note that we use the same notation for both the mean field value and the thermal average of
$\abs{\psi_i}^2$. It should be clear from the context which one is discussed. We also measure the
thermal average of the density as a function of position, $\langle\left|\psi_i(\bv
r)\right|^2\rangle$ to examine possible modulations in the density substrate. To monitor the thermal
fluctuations in the condensate densities, we compute their probability distribution,
$\mathcal{P}(\left|\psi_i\right|^2)$ by making a histogram of the field configurations at each
measuring step of the Monte-Carlo simulations.

In order to monitor the formation of the modulated ground state, we compute the phase
correlation function, defined by
\begin{equation}
  G_X(\bv r, \bv r^\prime) = \ev{e^{i\theta_{\bv r, X}}e^{-i\theta_{\bv r^\prime, X}}}
  \label{eq:sscorr}
\end{equation}
Here, $X$ may represent either component $1$, $2$, as well as the sum or difference of the two,
$\theta_1+\theta_2$ and $\theta_1-\theta_2$. We also calculate its fourier transform, the phase
structure function, defined by
\begin{equation}
  G_X(\bv q) = \frac{1}{V}\sum_{\bv r, \bv r^\prime}e^{i\bv q\cdot(\bv r-\bv r^\prime)}G_X(\bv r, \bv
  r^\prime).
  \label{eq:sssf}
\end{equation}
At large distances, $r$, the correlation function is expected to scale as $G_X(r)\sim r^{-\eta}$.
We measure this exponent by extracting the value of $G_X(\bv q)$ at a particular value, $\bv Q$,
which in turn defines the exponent $\eta_{\bv Q}$ as
\begin{equation}
  G_X(\bv Q)\sim L^{2-\eta}.
\end{equation}

\section{Mean-field theory}
\label{sec:mft}

Inter-component density interactions suppress the minority condensate at sufficiently
strong values of the coupling value. To get crude estimates for the interaction parameters needed
for this to occur, we start out by considering the model in the mean-field approximation. The full
fluctuation spectrum of the bosonic ordering fields will be considered in subsequent sections.
Here, we give the mean-field theory in a continuum model.

In order to account for the fact that the ground state generically is modulated in the presence
of SOC, we assume that the complex fields $\psi_i$ are given in terms of a mean
field value plus fluctuations, multiplied by a spatial plane wave modulation with momentum $\bv q$.
In general, we may use the ansatz\cite{Wang2010, Sedrakyan2012}
\begin{align}
  \psi_{1, \bv q} ={}& \sqrt{u_1 + \delta u_1}\exp{i(\phi_1+\delta\phi_1-\arg{\bv q} + \bv q\cdot\bv r)}\\
  \psi_{2, \bv q} ={}& \sqrt{u_2 + \delta u_2}\exp{i(\phi_2+\delta\phi_2+\bv q\cdot\bv r)},
\end{align}
where $\arg\bv q$ is the orientation of $\bv q$ with respect to some reference axis.  Specifically,
we follow previous work\cite{Wang2010, Sedrakyan2012} and assume that the ground state is either
modulated by a single wave vector (denoted $\Psi_0$), or by two oppositely aligned wave vectors
(denoted $\Psi_\pi$). That is
\begin{equation}
\Psi_0 = \begin{pmatrix}\psi_{1, \bv q}\\\psi_{2, \bv q}\end{pmatrix},
\end{equation}
and
\begin{align}
  \Psi_\pi ={}& \frac{1}{2}\begin{pmatrix}
  \psi_{1, \bv q}+\psi_{1, -\bv q}\\\psi_{2, \bv q}+\psi_{2, -\bv q}
  \end{pmatrix}\nonumber\\
  ={}&\begin{pmatrix}
  -\sqrt{u_1+\delta u_1}e^{i\phi_1+i\delta\phi_1-i\bar\theta}\sin\bv q\cdot\bv r\\
\sqrt{u_2+\delta u_2}e^{i\phi_2+i\delta\phi_2}\cos\bv q\cdot\bv r
\end{pmatrix},
\end{align}
where $\bar\theta$ is the average angle of $\bv q$ and $-\bv q$ with respect to the $x$-axis.
Here, the amplitudes, phases, and the wave-vectors are to be regarded as variational
parameters in the mean-field free energy of the modulated state.

Inserting these expression into \cref{eq:GL_modelbase}
and using the mean-field values only, we obtain the two free energy densities $f_0$ and $f_\pi$.
\begin{equation}
  f_0 = \frac{\left|\bv q\right|^2}{2}\left(u_1+u_2\right) -2\left|\bv q\right|\kappa
  \sqrt{u_1u_2}\sin(\phi_1-\phi_2)+V_0.
\label{mft_free_energy0}
\end{equation}
\begin{equation}
  f_\pi = \frac{\left|\bv q\right|^2}{4}\left(u_1+u_2\right) -\left|\bv q\right|\kappa
  \sqrt{u_1u_2}\cos(\phi_1-\phi_2)+V_\pi.
\label{mft_free_energypi}
\end{equation}
Here, the potentials $V_0$ and $V_\pi$ differ slightly due to numerical factors obtained when
integrating over space. They have the forms
\begin{align}
  V_0 = &-\alpha\Big[(1-\Delta)u_1+(1+\Delta)u_2\Big]\nonumber\\
        &+g\Big[(1-\gamma)u_1^2+(1+\gamma)u_2^2+2\lambda u_1u_2\Big]
\label{mft_free_energypot0}
\end{align}
and
\begin{align}
  V_\pi = &-\frac{\alpha}{2}\Big((1-\Delta)+u_1(1+\Delta)u_2\Big)\nonumber\\
          &+\frac{g}{8}\Big[3(1-\gamma)u_1^2+3(1+\gamma)u_2^2+2\lambda u_1u_2\Big]
\label{mft_free_energypotpi}
\end{align}
Note from \cref{mft_free_energy0,mft_free_energypi}, that in a modulated ground state the SOC
essentially acts as a phase locking on $\phi_1-\phi_2$ in a system with a uniform ground state. We
may minimize \cref{mft_free_energy0,mft_free_energypi} with respect to this phase difference,
assuming that $\left|q\right|\neq 0$ and $u_i\neq 0\;\forall\;i$, which yields a phase locking of
$\phi_1-\phi_2 = \pi/2$ for $f_0$ and $\phi_1-\phi_2 = 0$ for $f_\pi$. The angle $\arg\bv q$ in the
single $q$-vector case and the average angle $\bar\theta$ drops out of the equations, which reflects
the degeneracy of the single particle spectrum.

Considering the modulation vector present in \cref{mft_free_energy0,mft_free_energypi} as a variational parameter and
assuming $u_i\neq 0\; \forall \;i$, we find
\begin{equation}
  \left|\bv q\right| = \frac{2\kappa\sqrt{u_1u_2}}{u_1+u_2}
  \label{q-equations}
\end{equation}
in both cases. With this solution inserted into the free energy densities, they become
\begin{equation}
  f_0 =-\frac{2\kappa^2u_1u_2}{u_1+u_2}+V_0,
  \label{eq:feqins0}
\end{equation}
and
\begin{equation}
  f_\pi =-\frac{\kappa^2u_1u_2}{u_1+u_2}+V_\pi.
  \label{eq:feqinspi}
\end{equation}
\cref{eq:feqins0,eq:feqinspi} may in principle be solved for $u_1$ and $u_2$, but as they are cubic
the expressions for the solutions are unwieldy and not particularly illuminating. Instead, we
numerically minimize both free energy densities, and then determine the ground state for a given
parameter range by finding $\min(f_0, f_\pi)$. This gives the regions of the phase diagram where the
ground state is modulated by either one or two wave vectors. For the SOC to be effective, it is also
required that $u_1 u_2 \neq 0$.  For $u_1 u_2 = 0$, the model reverts to a single component
condensate, i.e. a "spinless" model where SOC cannot be operative.

\begin{figure}
\includegraphics[width=\columnwidth]{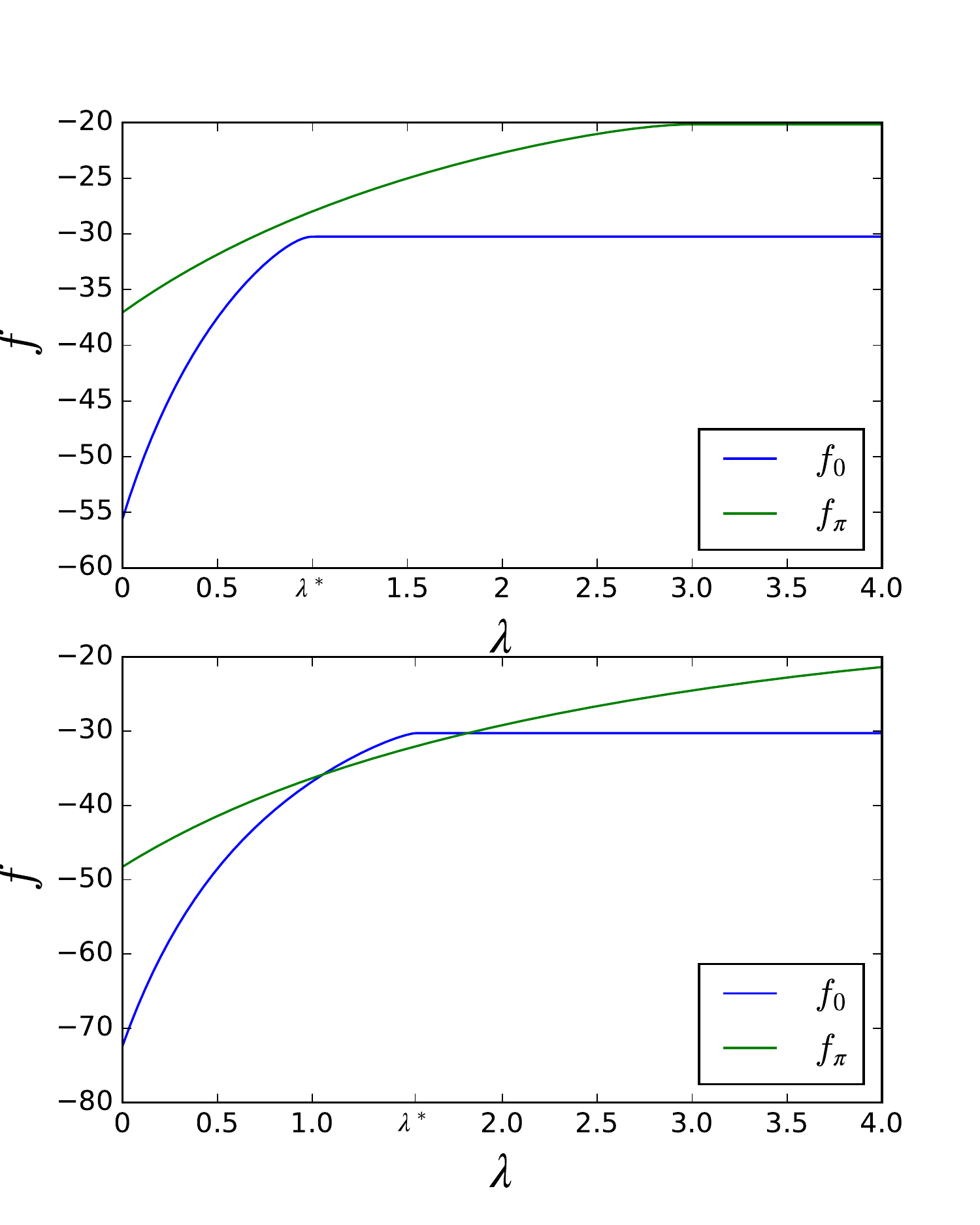}
\caption{Results for minimum values of $f_0$ and $f_\pi$ as a function of
$\lambda$ for two values of $\kappa$. Top panel: $\kappa=1$. Bottom panel:
$\kappa=2$. Note how $f_0$ ceases to be dependent on $\lambda$ for large
$\lambda$, at some value $\lambda^*$. Note also the discontinuity in the
derivative of $f_0$ at $\lambda = \lambda^*$.}
\label{fig:f0fpi}
\end{figure}

In \cref{fig:f0fpi}, we plot a few representative values of $f_0$ and $f_\pi$ as a function of
$\lambda$, for two  values of $\kappa$. For the lowest value of $\kappa$, it is seen that $f_0 <
f_\pi$ for all values of $\lambda$. Hence, a ground state modulated by two $q$-vectors is not found.
For a larger value of $\kappa$, $f_0 < f_\pi$ for low and high values of $\lambda$, while for
intermediate values of $\lambda$, $f_\pi < f_0$. Thus, for large enough $\kappa$ and intermediate values
of $\lambda$, there is the possibility of finding ground states modulated by two $q$-vectors.

Moreover, it is seen that for both values of $\kappa$, $f_0$ is independent of $\lambda$ when
$\lambda$ reaches some  value $\lambda= \lambda^*$. This happens at the value for which the minority
condensate ($u_1$ in this case) is completely suppressed. Furthermore, the second crossing of $f_0$
and $f_\pi$ always occurs at values of $\lambda > \lambda^*$. Therefore, for given $\kappa$ and with
increasing $\lambda$, the ground state modulated by two $q$-vectors always transitions into a
uniform ground state with one condensate completely suppressed.

Note also that $f_0$ increases more rapidly with $\lambda$ than $f_\pi$. This is due to difference
in the potentials $V_0$ and $V_\pi$, \cref{mft_free_energypot0,mft_free_energypotpi}. Therefore,
having two crossings of $f_0$ and $f_\pi$ as a function of $\lambda$ means that one of the crossing
points must always be to the right of the point where $f_0$ becomes $\lambda$-independent. Thus,
$f_0$ being minimal always transitions into $f_\pi$ being minimal II as $\lambda$ increases. There
will never be a transition from $f_\pi$ being minimal back to $f_0$ being minimal with increasing
$\lambda$.

\begin{figure}
\includegraphics[width=\columnwidth]{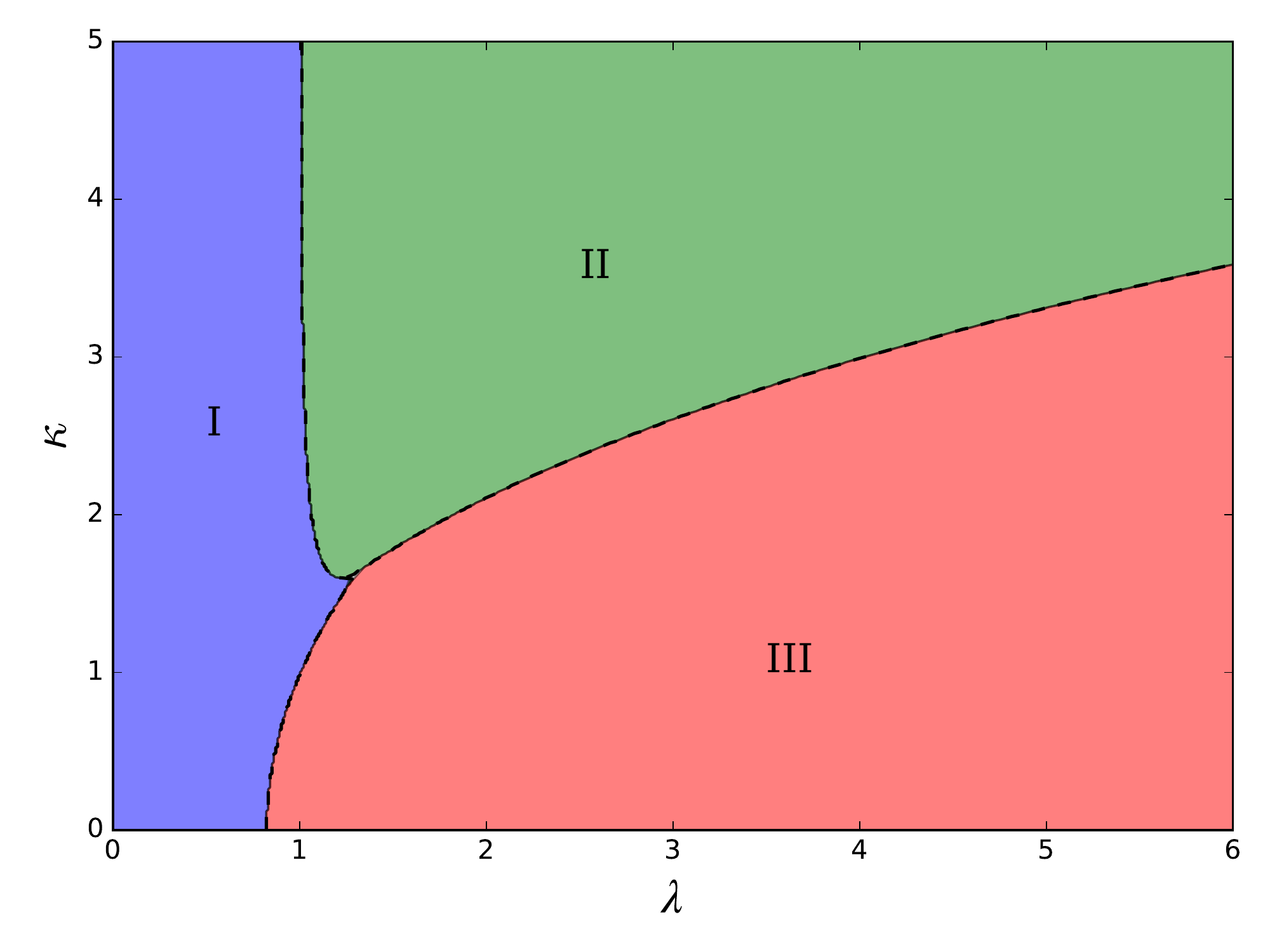}
\caption{Mean field phase diagram in the $\lambda$-$\kappa$ plane, with other
  parameters $\alpha=10.0$, $g=1.0$, $\Delta=0.1$, $\gamma=0.0$, $m=1.0$. Region I and III are the
  regions where both components exist and the effect of SOC is present, resulting in a ground state
  at finite momentum. In region I the ground state is modulated by a single wave-vector. In region
  III, the ground state is modulated by two oppositely directed wave vectors. Region II is the
  region where the inter-component interactions suppress the minority condensate, which results in a
  single-component condensate at zero momentum.
}
\label{fig:mfnum}
\end{figure}

This may be summarized as follows.  In \cref{fig:mfnum},  we show the results of numerically solving
\cref{eq:feqins0,eq:feqinspi} in the $\lambda-\kappa$ plane. Region I represents the area where the
single-$q$ modulated ground state is preferred, region III where the two-$q$ modulated ground state
is preferred, and region II is the area where $u_1 u_2 =0$ minimizes the free energy, making this
state a uniform, single-component state. The two lines separating I and II, and II and III are
located by the crossings of the free energies $f_0$ and $f_\pi$, and they therefore represent
first-order phase transitions at the mean-field level. The line separating region I and II is a
direct transition between a ground state modulated by one $q$-vector and a uniform ground state,
without an intermediate ground state modulated by two $q$-vectors. The location of this line is
therefore determined by the value of $\lambda$ where $f_0$ ceases to the dependent on $\lambda$,
while $f_\pi$ represents a higher-energy state which is irrelevant.  The order of this
phase-transition is determined by whether $\partial f_0/\partial \lambda$ is continuous or
discontinuous at $\lambda_c$. We have $\partial f_0/\partial \lambda \approx (\partial f_0/\partial
u_1) (\partial u_1/\partial \lambda)$. Using Eq. \ref{eq:feqins0}, we see that this is determined by
$\partial u_1/\partial \lambda$.  Since $u_1$ vanishes in a finite interval in $\lambda$, $\partial
u_1/\partial \lambda$ has to be discontinuous at $\lambda^*$, and hence so does $f_0$. The
transition line separating I and II is therefore also first order.

\section{Details of the Monte-Carlo simulations}
\label{sec:MCdetails}

The model is simulated using the Monte-Carlo algorithm with a simple restricted update scheme of
each physical variable, using Metropolis-Hastings\cite{Metropolis1953, Hastings1970} tests for
acceptance. The model is discretized on a rectangular lattice of size $L_x\times L_y$, with periodic
boundary conditions. Typically, $5\cdot10^6$ Monte-Carlo sweeps is used at each temperature step,
with an additional $5\cdot 10^5$ sweeps discarded for equilibration. One sweep consist of attempting
to update each physical variable on each lattice site once in succession. The proposed new value for
each variable is picked within a restricted region around the old value, where the size of the
region is chosen to allow for both high acceptance rates, and low autocorrelation times. To further
minimize auto-correlation times and increase simulation efficiency, we measure observables with a
period of $100$ Monte-Carlo sweeps. Pseudo-random numbers are generated with the Mersenne-Twister
algorithm\cite{Matsumoto1998}. During equilibration, time series of the internal energy is examined
for convergence, this ensures proper equilibration. To avoid metastable states, several simulations
with identical parameters, but differing initial seeds of the pseudo-random number generator are
performed to make sure they anneal to the same state. Measurements are post-processed using
multiple-histogram re-weighting\cite{Ferrenberg1989}, and error estimates are determined with the
Jackknife method\cite{Berg1992}.

The allowed range of amplitude fluctuations is determined during the equilibration procedure, by
first allowing it to fluctuate to a very large value ($\left|\psi_i\right|^2\sim 10$ was typically
used) and then reducing the value to include all values that had a non-zero probabilty of being
picked according to the measured probability distribution, $\mathcal{P}(\left|\psi_i\right|^2)$.

Unless otherwise stated, we fix $\alpha_0=10.0$, $g=1.0$, and $\gamma=0.0$.  The large value of
$\alpha_0=10.0$ is chosen to have sharp probability distributions of the amplitudes. Generally, a
square lattice of $L_x=L_y\equiv L=64$ is used in simulations, but system sizes of $L\in(16, 24, 32,
40, 48, 56, 64, 96, 128, 160, 192, 224, 256)$ are used for performing a finite size scaling
(FSS) analysis.

\section{Results of the Monte-Carlo simulations}
\label{sec:results}

In this section, we present Monte-Carlo simulations to corroborate and expand on the arguments given
in the previous sections. The model exhibits three different classes of BECs for different parameter
regimes. For strong inter-component interactions and zero to intermediate SOC, there will be only
one superfluid condensate present. With no SOC, but for intermediate inter-component interactions,
the model is a two-component coupled superfluid. Finally, for intermediate interactions and SOC, the
model is a two-component superfluid with a finite $q$-vector. This schematic picture shown in
\cref{fig:mfnum} is captured by a simple mean field argument, but we find it to be essentially
correct also when thermal fluctuations are taken into account in Monte-Carlo simulations. We also
examine the thermal phase transitions present in the cases of zero SOC and when the condensate is
modulated by a single $q$-vector.



\subsection{Kosterlitz-Thouless transition in the absence of spin-orbit coupling}
\label{sec:KTtrans}

When $\kappa=0$, the model represents a two-component BEC coupled by density-density interactions,
which may collapse to a single-component condensate for strong inter-component interactions. When
neglecting amplitude fluctuations (which of course decouples the condensates), the model reduces to
the XY-model. Here, the low-temperature phase is characterized by quasi long-range order of the
superfluid order parameter, where vortices and anti-vortices form bound pairs. As the temperature is
increased, the bound vortex-antivortex pairs unbind at a Kosterlitz-Thouless (KT) transition
\cite{Weber1988,Minnhagen2003}. As a check of simulations we indeed obtain that the two-component
model with amplitude fluctuations included belongs in the KT universality class by establishing that
the helicity modulus undergoes a discontinuous jump to zero as the system is heated from the
low-temperature state, with the value of the jump close to the predicted universal value. We
examine various values for the inter-component coupling $\lambda$, and find that the above remains
true for all the values of $\lambda$ we have considered.

\begin{figure}
\captionsetup[subfigure]{labelformat=empty}
\centering
\subfloat[]{\includegraphics[width=\columnwidth]{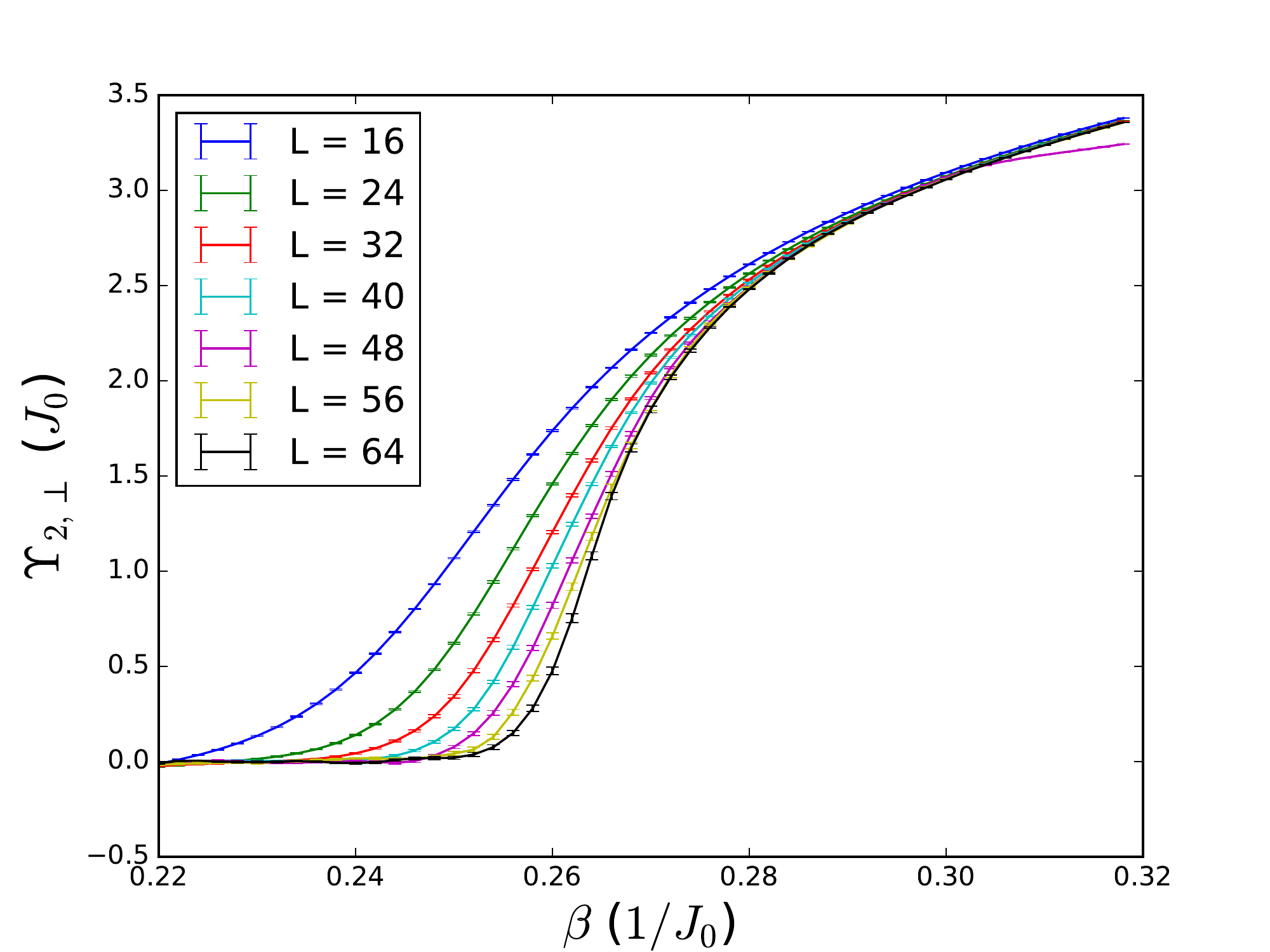}
\label{fig:KTheli}}\\
\subfloat[]{\includegraphics[width=\columnwidth]{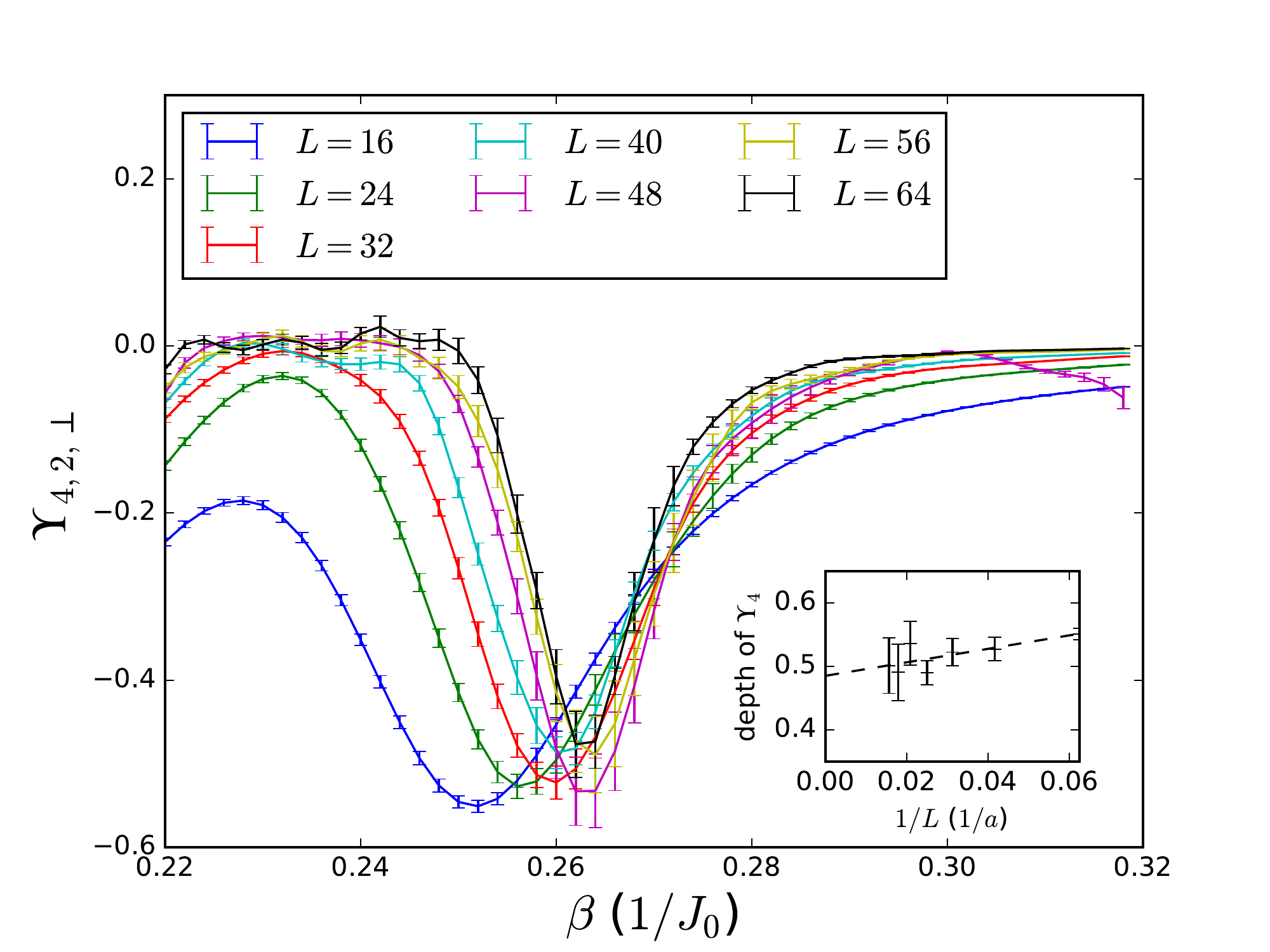}
\label{fig:KTmod4}}
\caption{Helicity modulus (top panel) and fourth order modulus (bottom panel) of component $2$ as a function of
  $\beta$ for several system sizes, with $\lambda=2.0$ and $\kappa=0.0$. The inset of the bottom
  figure shows the value of the dip in the fourth order modulus as a function of inverse system
  size. The dashed line is a linear extrapolation to the thermodynamic limit. At this value of the
inter-component coupling strength, the condensate density of component $1$ is extinguished, and hence
exhibits no KT-transition.}
\label{fig:KTtransition}
\end{figure}

\cref{fig:KTtransition} shows the helicity modulus and fourth order modulus of component $2$ for
system sizes $L\in(16, 24, 32, 40, 48, 56, 64)$ with inter-component coupling strength $\lambda=2.0$.
The inset shows the depth of the dip in the fourth order modulus as a function of inverse linear
system size.  By fitting the helicity modulus to \cref{eq:helifit} we determine the discontinuous
jump to be $\Upsilon(\infty)\beta_c=0.650(1)$ at $\beta_c=0.282$. Extrapolation of the value of the
negative dip to $1/L=0$ gives a finite value of $0.49(1)$. This is clear evidence for a
discontinuous jump in the helicity modulus, placing the transition in the Kosterlitz-Thouless
universality class.

Similar results are obtained for values of $\lambda\in(0.0, 0.25, 0.5, 0.75, 1.25, 1.5, 1.75, 2.0)$
as shown in \cref{tab:KTtransitions}. For the values of $\lambda$ where both condensates persist,
transitions of KT type is observed in both components, at different critical couplings. In all cases
considered, the value of the minimum in $\Upsilon_4$ converges to a nonzero value. This demonstrates
that there is a discontinuous jump in the helicity modulus, regardless of the value of the
inter-component interaction strength, and whether or not the minority condensate is depleted.
Additionally, the value of the discontinuous jump varies weakly with $\lambda$, and is close to the
universal value of $2/\pi$. This indicates that fluctuations in the condensate amplitude only have a
minor effect on the details of the transition. None of the obtained jumps are within the prediction
$2/\pi$, with error estimates, but most are close. Moreover, the fitting routine was sensitive to
the system sizes that were included. Both effects may have been caused by the inclusion of amplitude
fluctuations. Also note that the critical temperature and depth of the dip varies very weakly with
$\lambda$, as long as $\lambda\geq1.0$. This is very reasonable, as the model is effectively a
single component condensate in this regime, so varying the inter-component interaction strength
should have little to no effect.

\begin{table*}
\setlength{\tabcolsep}{7pt}
\centering
\caption{Summary of the results obtained when searching for the KT-transition. Each row shows, for
  both components,  the critical inverse temperature at which the best fit to \cref{eq:helifit},
  $\beta_c$, the size of the jump at this inverse temperature, $\Upsilon_\infty\beta_c$, as well as
  the extrapolation of the value of the minimum in the fourth order modulus to $1/L=0$. When
$\lambda\geq1.0$, the density of component $1$ has been completely depleted, and there is no phase
transition in this sector, as signified by the entries marked N/A.}
\begin{tabular}{c c c c c c c c}
  \hline\hline
  & \multicolumn{3}{c}{Component $1$} & & \multicolumn{3}{c}{Component $2$}\\
  \cline{2-4}\cline{6-8}
  $\lambda$ & $\beta_c$  & $\Upsilon_\infty\beta_c$   & value of minimum in
  $\Upsilon_4$  & & $\beta_c$  & $\Upsilon_\infty\beta_c$   & value of minimum in
  $\Upsilon_4$ \\
  \hline
  0.00 & 0.280 & 0.617(1) & 0.56(3)   && 0.226 & 0.642(2) & 0.58(7) \\
  0.25 & 0.391 & 0.609(1) & 0.367(8)  && 0.249 & 0.5(3)   & 0.67(3) \\
  0.50 & 0.605 & 0.595(1) & 0.239(9)  && 0.284 & 0.625(1) & 0.49(3) \\
  0.75 & 2.24  & 0.58(1)  & 0.068(4)  && 0.290 & 0.627(1) & 0.50(2) \\
  1.00 & N/A    & N/A      & N/A      && 0.292 & 0.662(1) & 0.48(2) \\
  1.25 & N/A    & N/A      & N/A      && 0.290 & 0.667(1) & 0.46(3) \\
  1.50 & N/A    & N/A      & N/A      && 0.290 & 0.703(1) & 0.50(1) \\
  1.75 & N/A    & N/A      & N/A      && 0.284 & 0.653(1) & 0.61(5) \\
  2.00 & N/A    & N/A      & N/A      && 0.282 & 0.650(1) & 0.49(1) \\
  \hline \hline
\end{tabular}
\label{tab:KTtransitions}
\end{table*}

Finally, we remark that the fit of the discontinuous jump and the determination of the depth of the
dip in the fourth order modulus are two independent methods for detecting a KT-transition. As both
methods give good results consistent with the KT-prediction, we are confident in claiming that the
two-component imbalanced BEC without SOC has one or two transitions, depending on the value of the
inter-component coupling strength, in the KT universality class.{ However, pinning down a
KT-transition with great confidence is notoriously difficult. In particular, \cref{eq:helifit}
involves slowly decaying corrections that are suppressed only logarithmically. Several
works\cite{Stiansen2012,Herland2012} has utilized this particular method on various models 
with success, and methods to overcome the slowly decaying corrections exist\cite{Ceccarelli2013}. 
A detailed study of this is not the main focus of the present paper. We limit ourselves to 
noting that our results are consistent with a KT-transition, as is expected for the model 
in the absence of SOC.}

\subsection{Spin-orbit induced modulated ground states}
\label{sec:SOCGS}

Preliminary arguments based on the non-interacting energy spectrum and mean field calculations
suggest that the ground state of the spin-orbit coupled BEC resides at either one or two finite
$q$-vectors.  In order to confirm this, Monte-Carlo simulations of the full lattice model,
\cref{eq:Hlattice,eq:Hklattice,eq:HVlattice,eq:HSOlattice}, were performed in parameter regions
corresponding to region I and III in the phase diagram of \cref{fig:mfnum}.

\subsubsection{Single $q$-vector}

\begin{figure}
\centering
\includegraphics[width=\columnwidth]{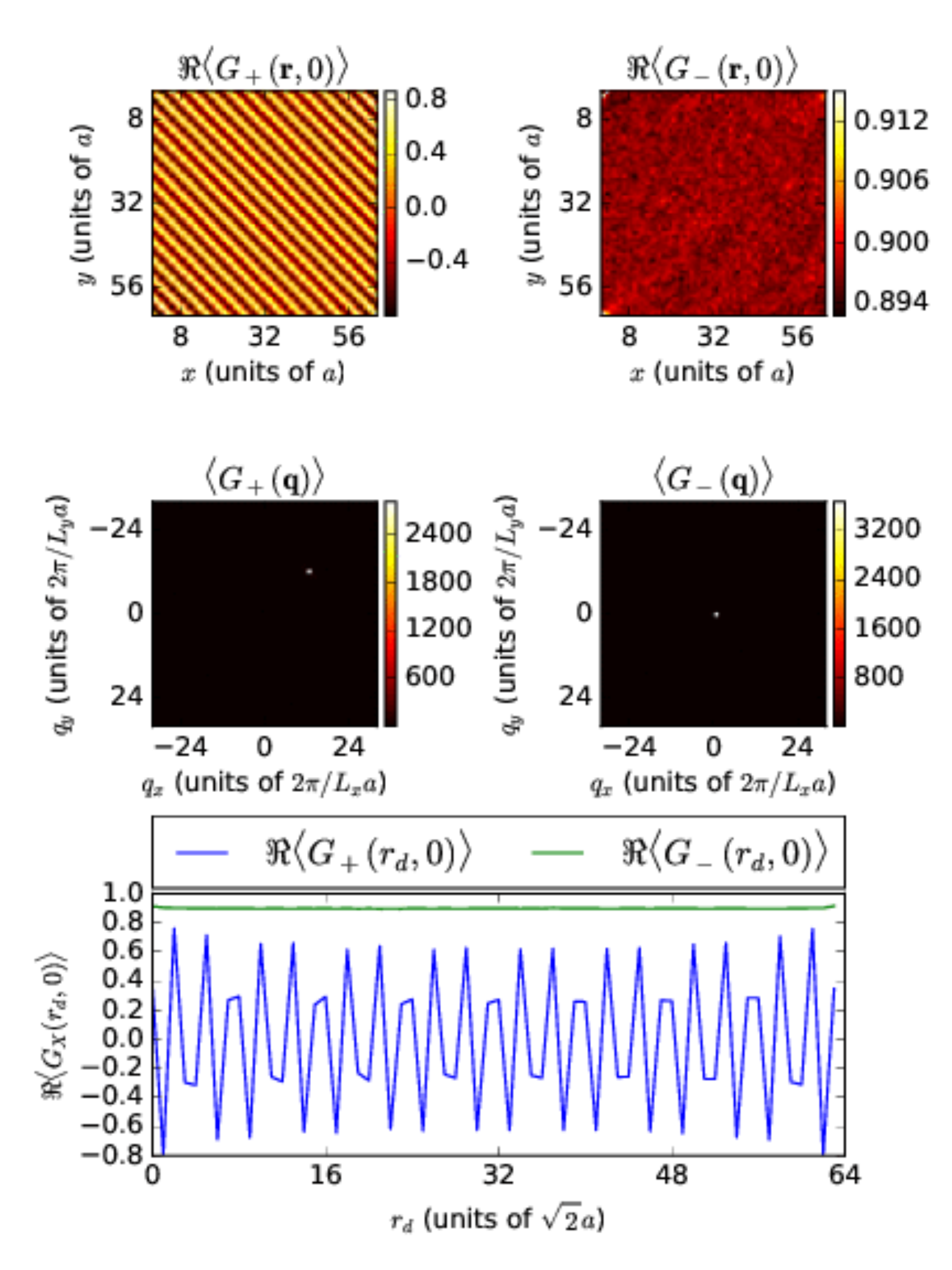}
\caption{The real part of the phase correlation function \cref{eq:sscorr} in real (top row)
and reciprocal (middle row) space of the phase sum (left column) and phase difference (right
column), at parameters $\kappa=1.0$, $\lambda=0.0$ and $\beta=1.0$. The bottom panel shows a
real-space cut along the diagonal perpendicular to the stripes, $r_d$, of both the phase sum and
phase difference correlation functions. The effect of the SOC is manifest in the phase sum, which
is modulated by a wave-vector, $\bv Q$. The phase difference, exhibits no modulations in
the spatial correlation. We have removed the reference point $\bv r = 0$ from the real
space plots to improve visibility of the correlations.}
\label{fig:spincorrmodulation}
\end{figure}

To observe the predicted modulated state where a single $q$-vector is present, we perform
simulations of the lattice model at $\kappa=1.0$ and $\lambda=0.0$.  \cref{fig:spincorrmodulation}
shows the real parts of the phase correlation function, \cref{eq:sscorr}, and the structure
factors of the phase sum and phase difference variable in the low temperature phase, when the
inverse temperature is $\beta=1.5$. The phase correlation function \cref{eq:sscorr} for the phase
sum composite variable is modulated with a single $q$-vector along the diagonal. The phase
difference composite variable shows no modulation. It is, however, highly correlated, which is a
result of the effective Josephson locking. This is in accord with expectations based on the
London-approximation, where amplitudes are frozen, see \cref{eq:HSOLondon}.  The London case,
with non-modulated amplitudes, suffices to describe the situation with relatively small values of
intercomponent density-density interactions, where amplitudes are constant throughout the system.
The SOC-term tends to lock $\theta_1-\theta_2$ at constant value, since the strength of the SOC-term
effectively is constant due to the constant values of the amplitudes, while SOC induces a gradient
in  $\theta_1+\theta_2$.  The $\theta_1+\theta_2$-modulations therefore originate with SOC-coupling.

\begin{figure}
\centering
\includegraphics[width=\columnwidth]{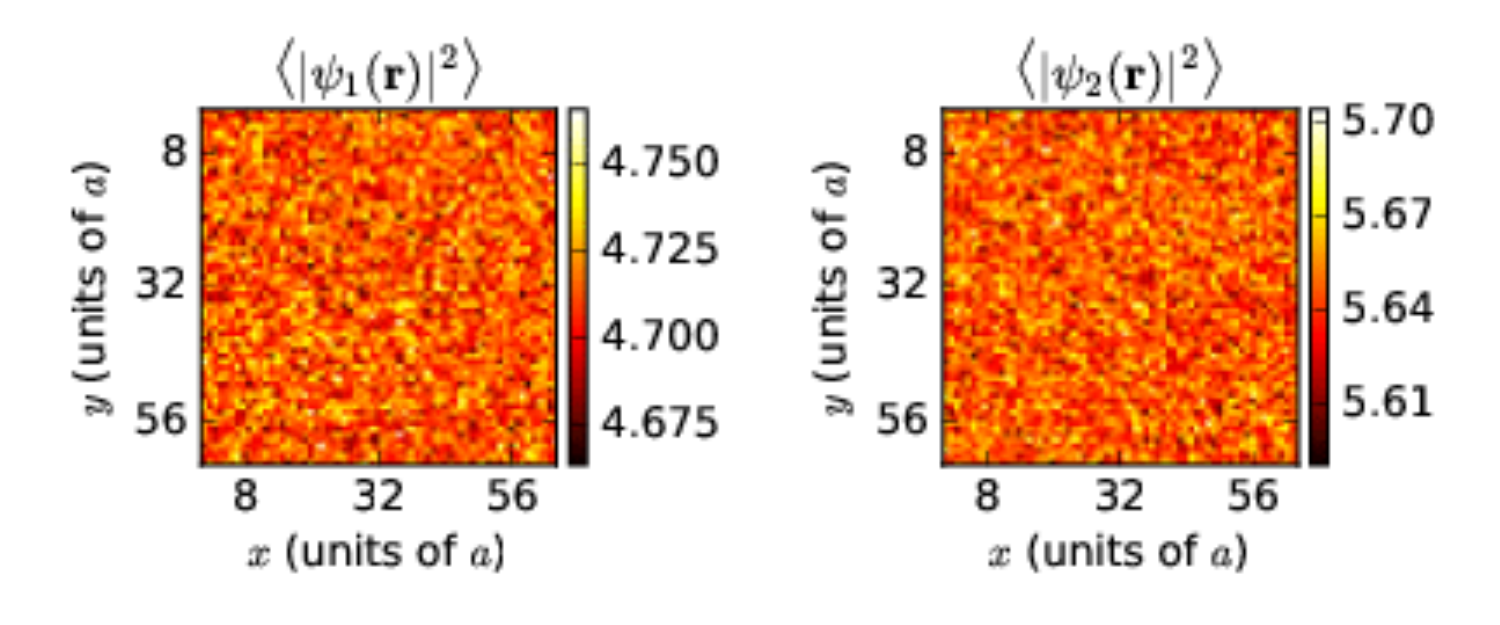}
\caption{Thermal amplitude averages in real space for component 1 (left panel) and 2
(right panel), at parameters $\kappa=1.0$, $\lambda=0.0$ and $\beta=1.0$. There is 
only minor spatial fluctuations around the
average, $u_i$, in each individual component.}
\label{fig:amps}
\end{figure}

In these simulations, the amplitudes are also allowed to fluctuate. The real-space amplitude
plots shown in \cref{fig:amps}, show that the spatial amplitude fluctuations are small. In
this regime the potential does not favor large density differences between the two components,
and there is no phase separation. The state we observe is the same as was found in
Refs. \onlinecite{Cole2012, Toniolo2014}, where a single minimum in the non-interacting
spectrum is populated for $\lambda<1$.

\subsubsection{Double $q$-vector}

\begin{figure}
\centering
\includegraphics[width=\columnwidth]{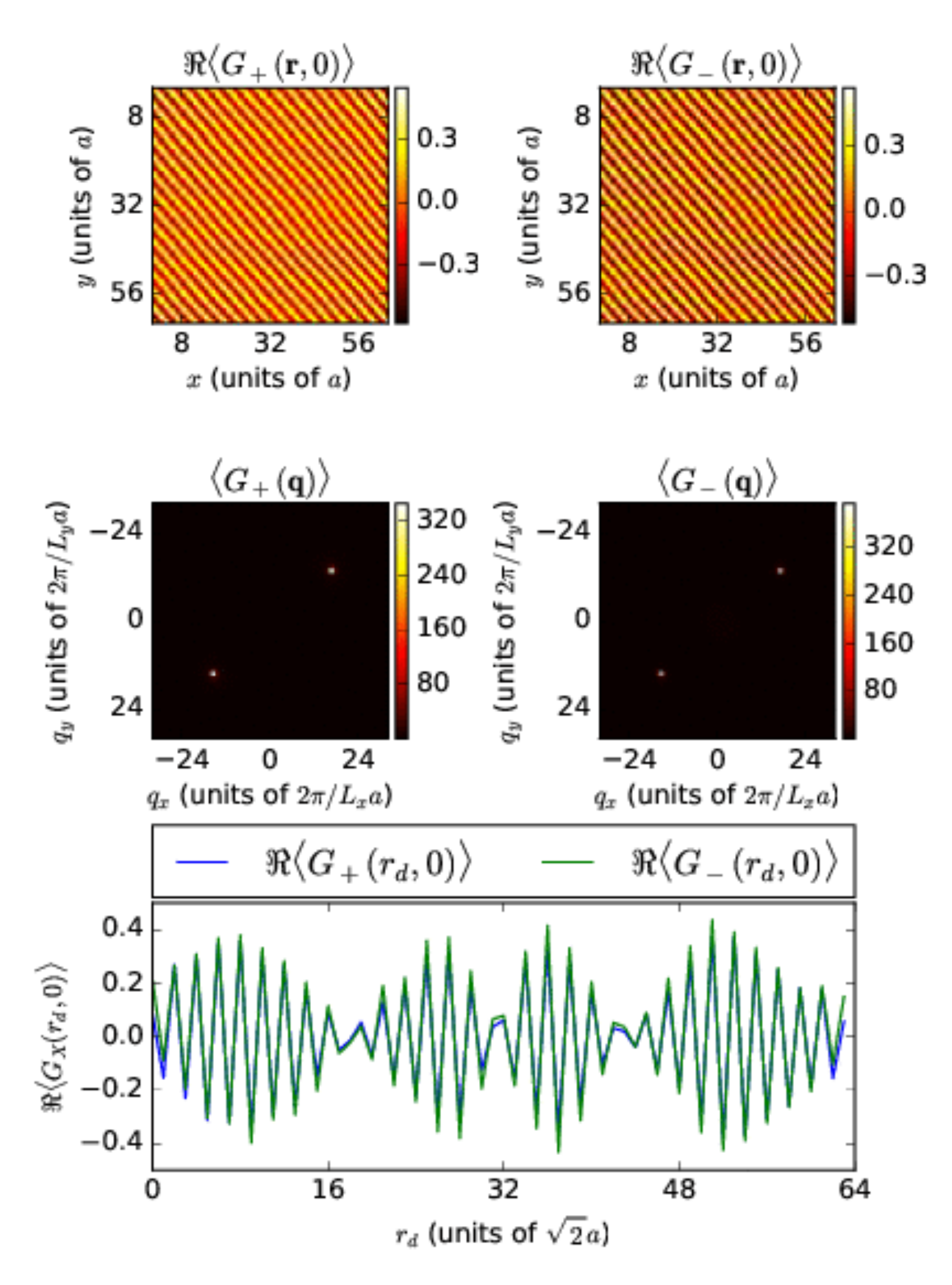}
\caption{Real part of phase correlation function \cref{eq:sscorr} in real (top row) and
  reciprocal (middle row) space of the phase sum (left column) and phase difference (right column),
  at parameters $\beta=1.0$, $\lambda=1.2$ and $\kappa=1.7$. In the bottom panel, we also show a
  real-space cut along the diagonal perpendicular to the stripes, $r_d$, of both correlation
  functions. It is shown that both the phase sum and the phase difference are modulated by two
  oppositely aligned wave-vectors, $\pm\bv Q$, with equal magnitude. We have removed the reference
  point $\bv r = 0$ from the real space plots to improve
visibility of the correlations.}
\label{fig:spincorrmodulationk2}
\end{figure}

The ground state modulated by two oppositely directed $q$-vectors only occurs, in mean field,
at sufficiently high values of both $\kappa$ and $\lambda$. In order to observe this state, we
perform simulations at $\kappa=1.7$ and $\lambda=1.2$, with $\beta=1.0$, inside region III of \cref{fig:mfnum}.
In \cref{fig:spincorrmodulationk2} we show Monte-Carlo calculations of the correlation function
of the phase sum and difference, in both real and reciprocal space. As in the single-$q$ vector
case, the phase-sum correlation is modulated, although now with a larger $\left|\bv q\right|$.
The increase of the length of the $q$-vector directly reflects the larger value of the
SOC strength.

\begin{figure}
\centering
\includegraphics[width=\columnwidth]{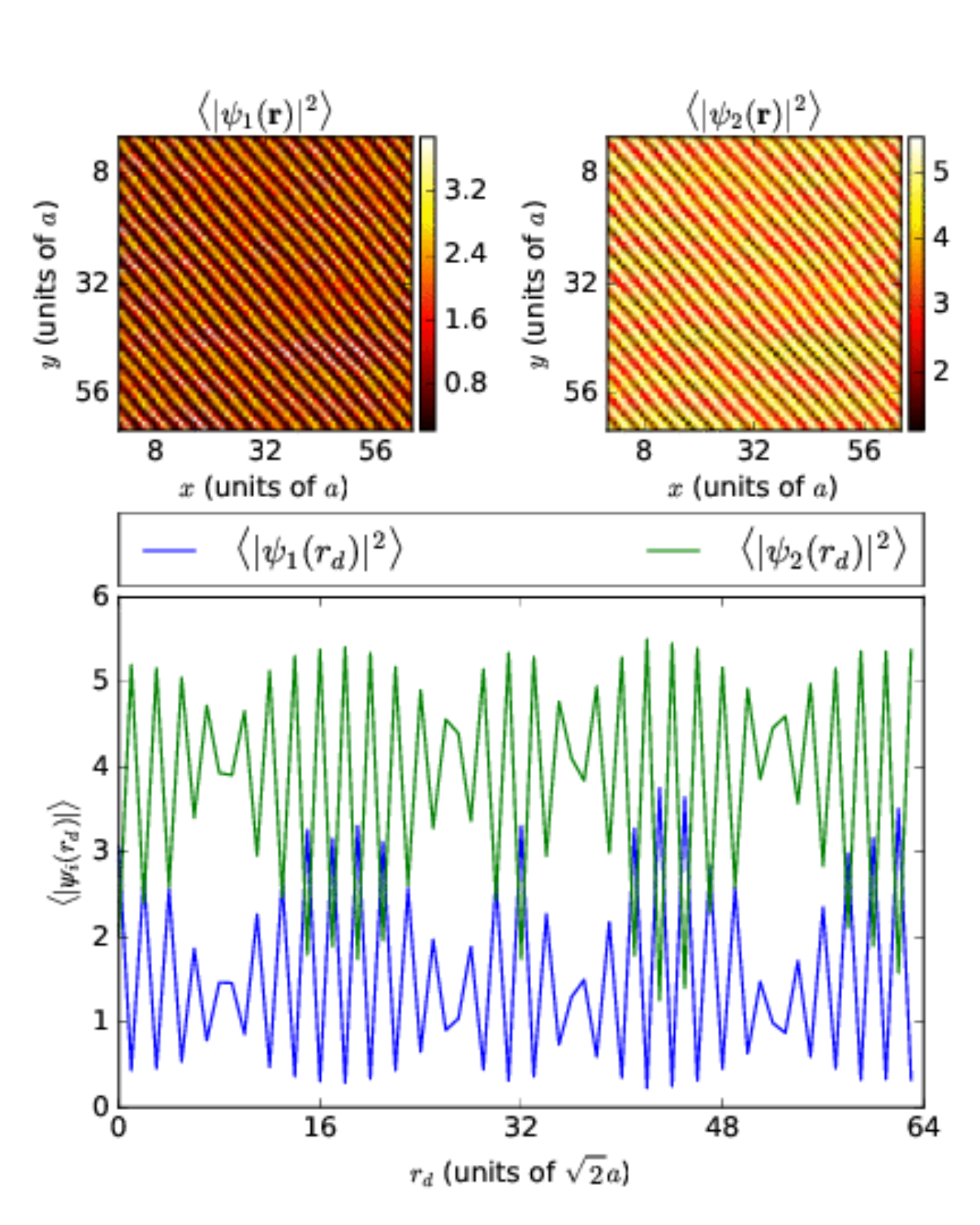}
\caption{Thermal amplitude averages in real space for component 1 (top row, left panel) and 2 (top,
  row, right panel), at parameters $\beta=1.0$, $\lambda=1.2$ and $\kappa=1.7$. The bottom panel shows
  a cut of the amplitude averages along the diagonal perpendicular to the stripe modulations, $r_d$. Both amplitudes are
modulated in this region of parameter space, but around different mean values because of the density
imbalance. Furthermore, the amplitude of component $1$ is staggered compared to component $2$. This
minimizes the potential energy from the inter-component density-density interaction, while still
minimizing the SOC interaction energy.}
\label{fig:ampsk2}
\end{figure}

Another important difference between the double-$q$ vector state compared to the single-$q$ vector
state is shown in \cref{fig:ampsk2}, which shows the thermal averages of the amplitudes. In this
case, the amplitudes are also modulated. Furthermore, the amplitudes of the two components are
staggered, when component $1$ has a large amplitude, component $2$ has a low amplitude, and vice
versa. This is further exemplified in the bottom panel of \cref{fig:ampsk2}, where we show a cut
along the diagonal perpendicular to the stripes in the amplitude densities. Here it is clearly seen
that the two amplitude variations are mirror images of each other, only shifted relative to each
other by the difference in the average amplitudes due to the component imbalance.

Unlike the single-$q$ vector case, the phase-difference correlation is also modulated.
This may now be understood as follows. The system is
in a parameter-regime where $\lambda$ is large enough to induce staggering of
the amplitudes of the condensates, in order to minimize energy. The London-approximation,
\cref{eq:HSOLondon}, therefore no longer suffices to describe the system, and
we revert to \cref{eq:HSOlattice}. It is the term with
the minus-sign in $H_{SO}$ that leads to the frustration of $\theta_1-\theta_1$.
Were this sign to be reversed, we would have had $\theta_1-\theta_2 = 0$.
Since the amplitudes are modulated, so are the gradients of the amplitudes, and so is
therefore the strength of the frustration in the phase-difference. This difference
is therefore itself modulated. The modulation of $\theta_1-\theta_1$ therefore
originates with the modulation of amplitudes, which is a consequence of strong
inter-component density-density interactions. Recall from above that the
modulation of $\theta_1+\theta_1$ originates with SOC.

\subsection{Interaction-induced destruction of modulated ground states}

The mean field calculations presented in \cref{sec:mft} predict a breakdown of the modulated ground
state shown in \cref{fig:spincorrmodulation} when the inter-component interaction parameter,
$\lambda$, reaches the threshold shown in \cref{fig:mfnum}, provided  $\kappa \lesssim 1.5$.  Above
this threshold, the condensate transitions from a single-$q$ condensate into a condensate modulated
by two opposite wave vectors. For $\gamma=0$, and $\Delta>0$, which we consider here, component $1$
is the minority component that collapses. The mechanism for the collapse is that inter-component
interactions drive the minority component to zero to eliminate the interaction energy. When the
model collapses to an effective one component model there will no effects of the SOC, as the
$q$-vectors of the modulation induced by it are proportional to $u_1u_2$, at the mean field level.

\begin{figure}
\centering
\includegraphics[width=\columnwidth]{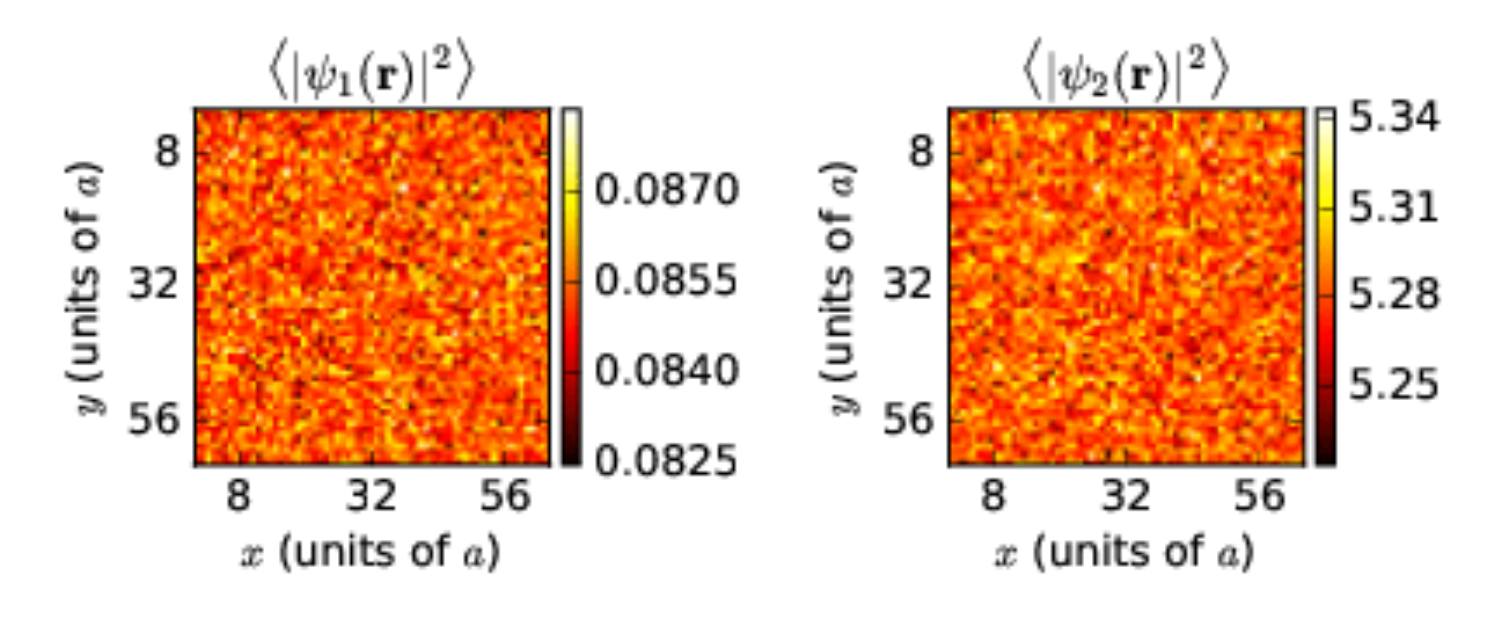}
\caption{Thermal amplitude averages in real space for component 1 (top row, left panel) and 2 (top,
  row, right panel), at parameters $\kappa=1.0$, $\lambda=2.0$ and $\beta=1.0$.}
\label{fig:ampsk1l2}
\end{figure}

To show this suppression, we compute the thermal amplitude averages of both components in the low
temperature phase, shown in \cref{fig:ampsk1l2}, when $\beta=1.0$, $\kappa=1.0$ and $\lambda=2.0$.
That is, every parameter is identical to what is shown in \cref{fig:spincorrmodulation,fig:amps},
except the inter-component interaction is increased above the critical value given by the mean field
calculations. It is evident that both amplitudes are now again unmodulated, but the amplitude of
component $1$ has been almost completely depleted. Its small finite value is only a remnant of
the thermal fluctuations included in the simulations.

\begin{figure}
\centering
\includegraphics[width=\columnwidth]{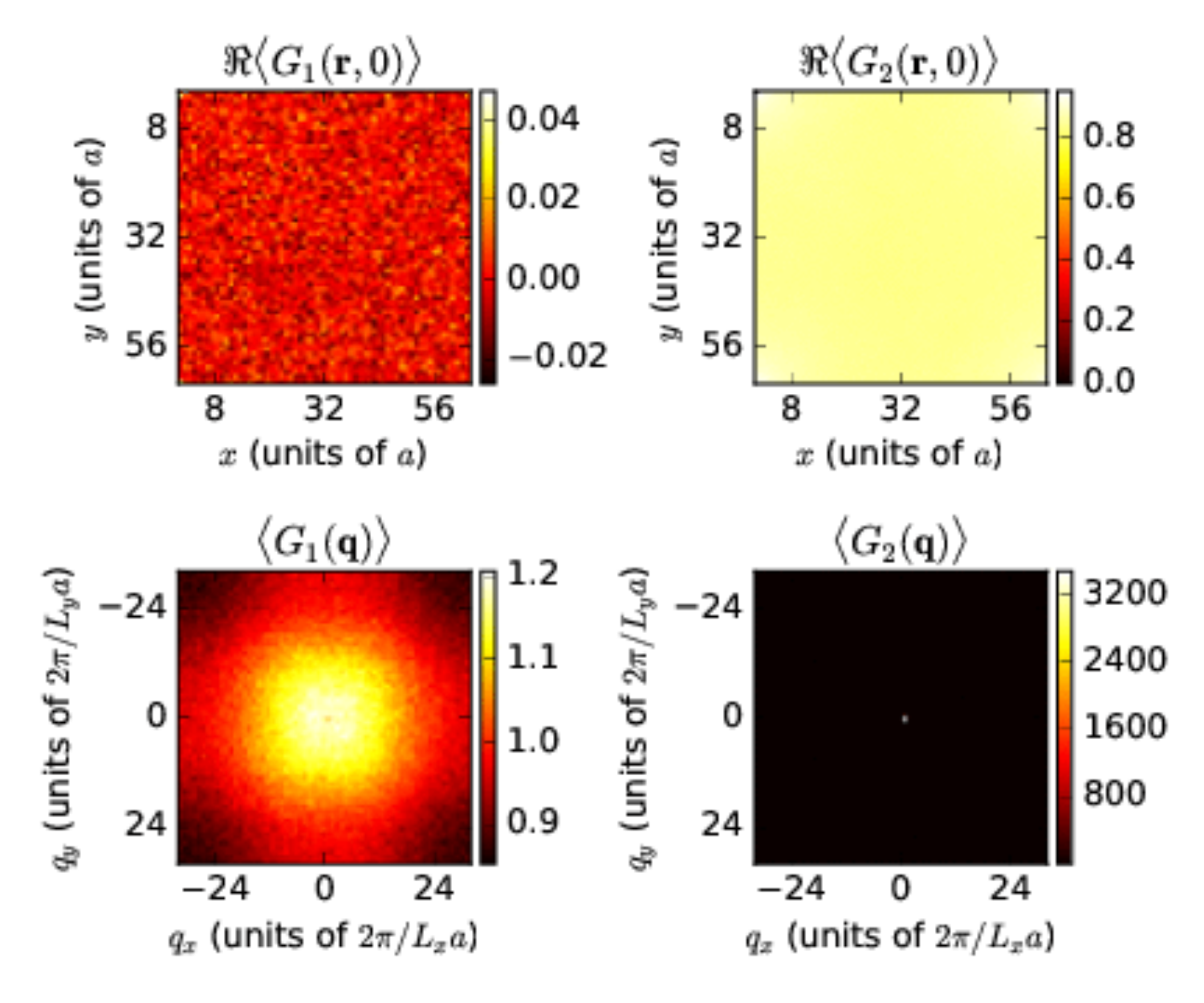}
\caption{Real part of the phase correlation function \cref{eq:sscorr}  in real
space (top row) and reciprocal space(bottom row) for the phase of component 1(left column)
and component 2(right column), at parameters $\kappa=1.0$, $\lambda=2.0$
and $\beta=1.0$. We have removed the reference point $\bv r = 0$ from the real
space plots to improve visibility of the correlations.}
\label{fig:spincorrnomodulation}
\end{figure}

To further explore the effect of the depletion, we compute the phase correlation function
\cref{eq:sscorr} and its Fourier transform, \cref{eq:sscorr} and \cref{eq:sssf}.
\cref{fig:spincorrnomodulation} shows the real parts of both the phase correlation function
\cref{eq:sscorr}, and structure factor of both individual componentsThere are no modulations in the
either of the phase correlation function \cref{eq:sscorr}, and both structure factors are isotropic.
However, while the phase of component $1$ is completely uncorrelated, the phase of component $2$ is
strongly correlated. The reasons for this is that: \textit{i)} the condensate amplitude of component
$1$ has been completely depleted, leaving the phase of this component completely uncorrelated at all
temperatures, and \textit{ii)} the non-suppressed condensate has entered a low-temperature
superfluid state, akin to what we observe for $\kappa=0$, even though we still have a finite SOC,
however ineffective.

\begin{figure}
\centering
\includegraphics[width=\columnwidth]{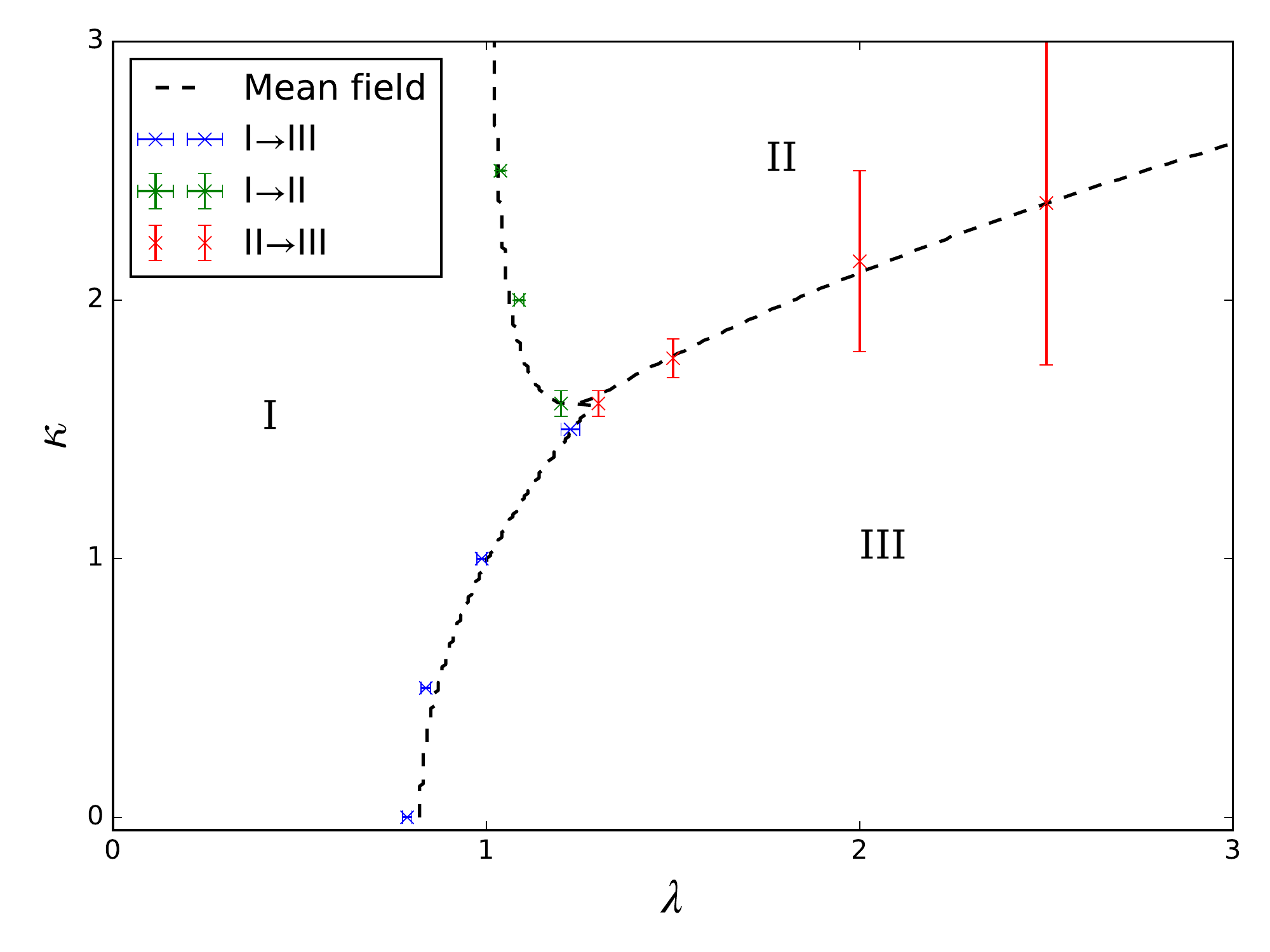}
\caption{Phase diagram obtained from numerical Monte-Carlo simulations compared to mean field
predictions. The points with error bars correspond to observed transition points, blue points
correspond to the transition from region I to region II, green points the transition from region I
to region III, and red points the transition from region II to region III. The dashed lines are the
corresponding transition lines obtained from mean field calculations, shown in \cref{fig:mfnum}}
\label{fig:mfmcnum}
\end{figure}

\cref{fig:mfmcnum} summarizes the results obtained in the Monte-Carlo simulations, showing an
overview of the different ground states obtained at slow annealing from a random initial state at
high temperature down to $\beta=6.0$, for different values of $(\kappa,\lambda)$.  The size of
region I was largely unaffected. For intermediate values of $\kappa$ and sufficiently large values
of $\lambda$, we observe that the spin-orbit induced modulations of both the amplitudes and the
phases are pinned to the crystal axes of the numerical lattice. This is represented by the large
error bars of the red points denoting the transition from region II to region III obtained from the
Monte-Carlo simulations. We determine these particular error bars by finding the upper and lower
limits in $\kappa$ where we can confidently observe a pure double $q$-vector condensate, or a pure
single component condensate. That aside, the mean-field and MC calculations correspond remarkably
well, even close to the area where the three transition lines meet.

\subsection{Thermal disordering of single-$q$ modulated state}

Thermal fluctuations of the superfluid phases are also expected to disorder the modulated ground
state pattern induced by the SOC. The modulation which appears in region I at low temperatures is
characterized by modulated superfluid order, or superfluid order with a texture. The temperature
driven disordering of this modulated superfluid state is expected to lie in the KT-universality
class. In order to examine the thermal phase transition from the low temperature phase of region I
into the high temperature phase, we perform simulations of the full Hamiltonian as written in
\cref{eq:Hlattice} and in the London limit. The London limit is employed here as it is the minimal
model which captures the effect of the SOC. As discussed in section \cref{sec:SOCGS}, in region I
where the condensate is only modulated by a single $\bv q$-vector, we find that the amplitudes are
essentially uniform. Hence, the amplitude fluctuations are largely irrelevant for this phase, and we
may therefore employ the London limit.  The London limit is taken by fixing $\left|\psi_{\bv r,
i}\right|=1\;\forall\;\bv r,i$, which simplifies the Hamiltonian greatly.


\begin{figure}
\captionsetup[subfigure]{labelformat=empty}
\centering
\subfloat[]{\includegraphics[width=\columnwidth]{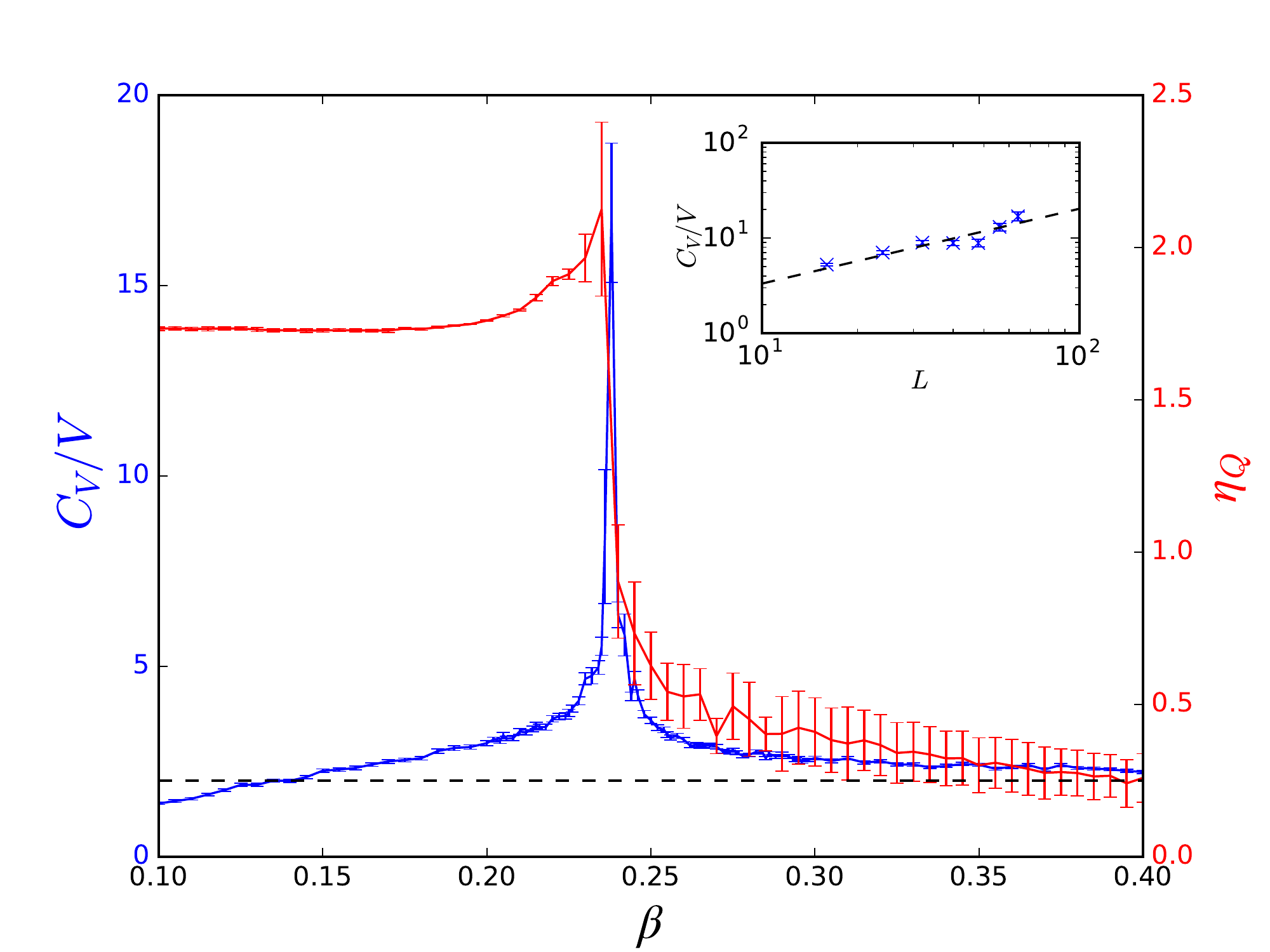}
\label{fig:cvetanotrapnoLondon}}\\
\subfloat[]{\includegraphics[width=\columnwidth]{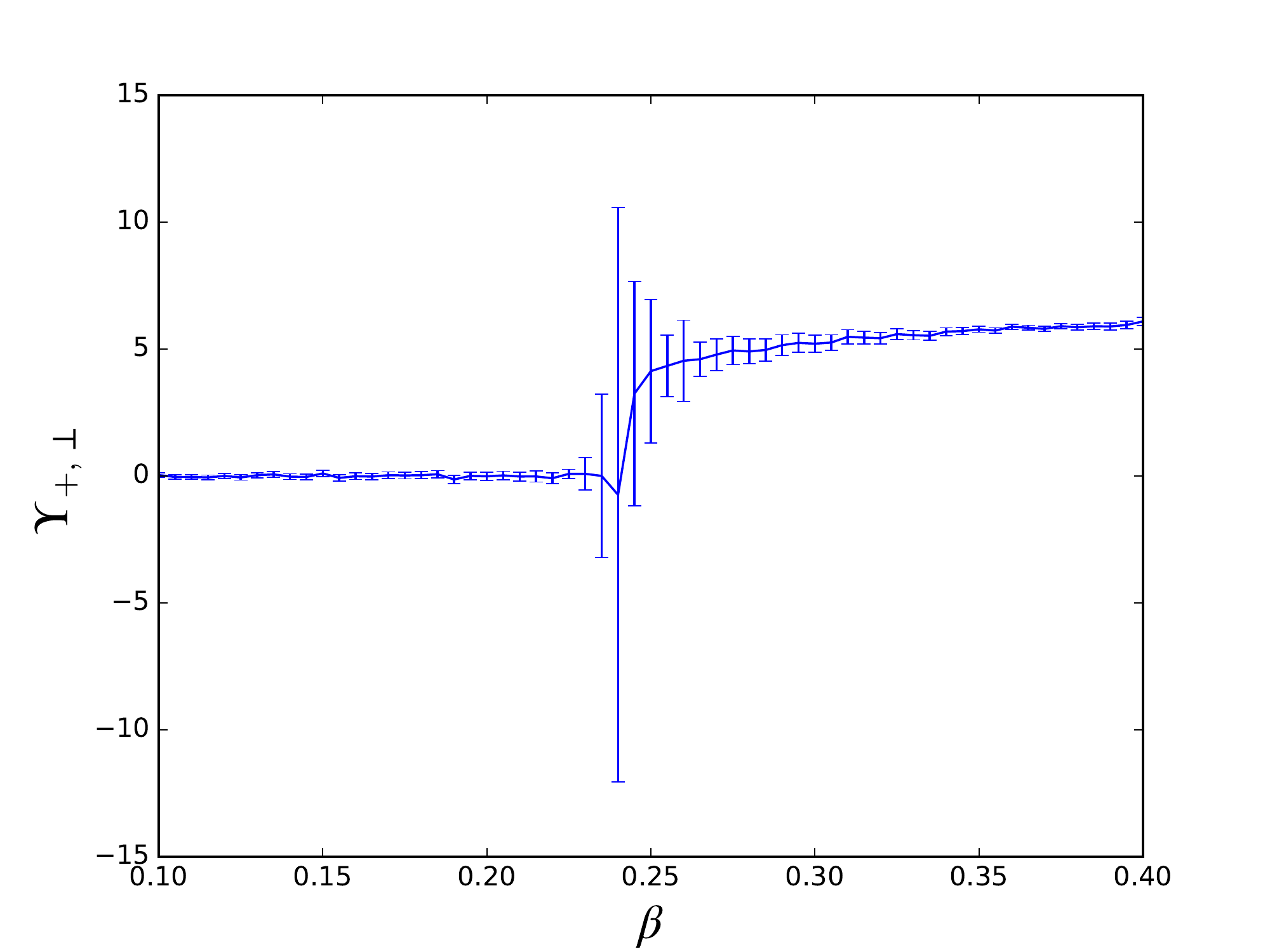}
\label{fig:helinotrapnoLondon}}
\caption{Phase-sum structure function at the first Bragg peak, $\bv Q$, as a function of $\beta$ for
  system sizes $L\in(16, 24, 32, 40, 48, 56, 64)$ as well as specific heat $C_V/V$ for $L=64$ (top),
  and helicity modulus of the phase-sum variable, $\Upsilon_{+, \perp}$ (bottom), at $\kappa=1.0$ in
  the London limit. The inset of the top panel show the scaling of the peak of the specific heat
  curves for the same system sizes used in the structure function scaling. Note how the drop in the
  exponent $\eta_{\bv Q}$ as well as the jump in the helicity modulus both coincide with the sharp
  peak in the specific heat.}
\label{fig:notrapnoLondon}
\end{figure}

\begin{figure}
\captionsetup[subfigure]{labelformat=empty}
\centering
\subfloat[]{\includegraphics[width=\columnwidth]{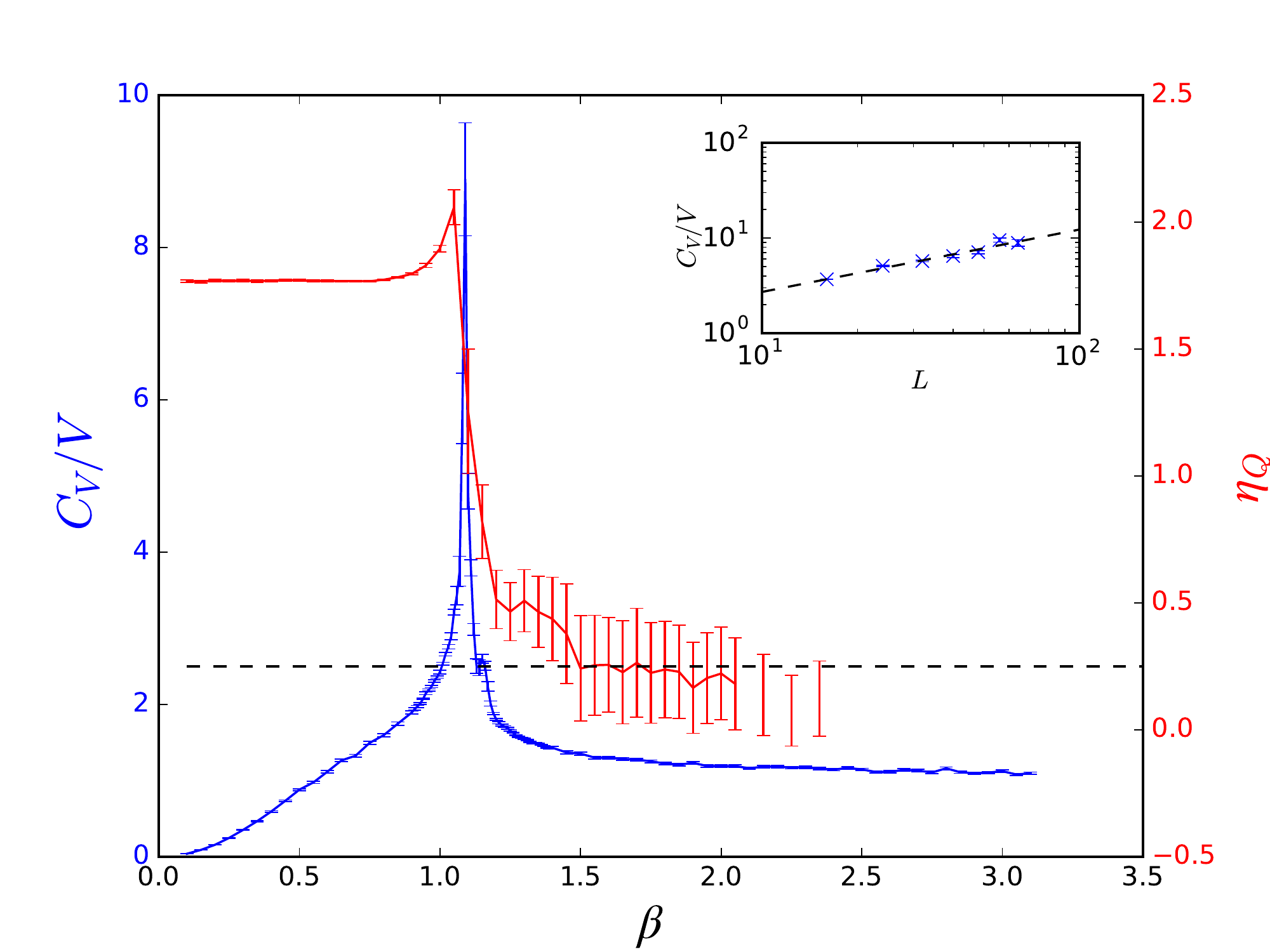}
\label{fig:cvetanotrapLondon}}\\
\subfloat[]{\includegraphics[width=\columnwidth]{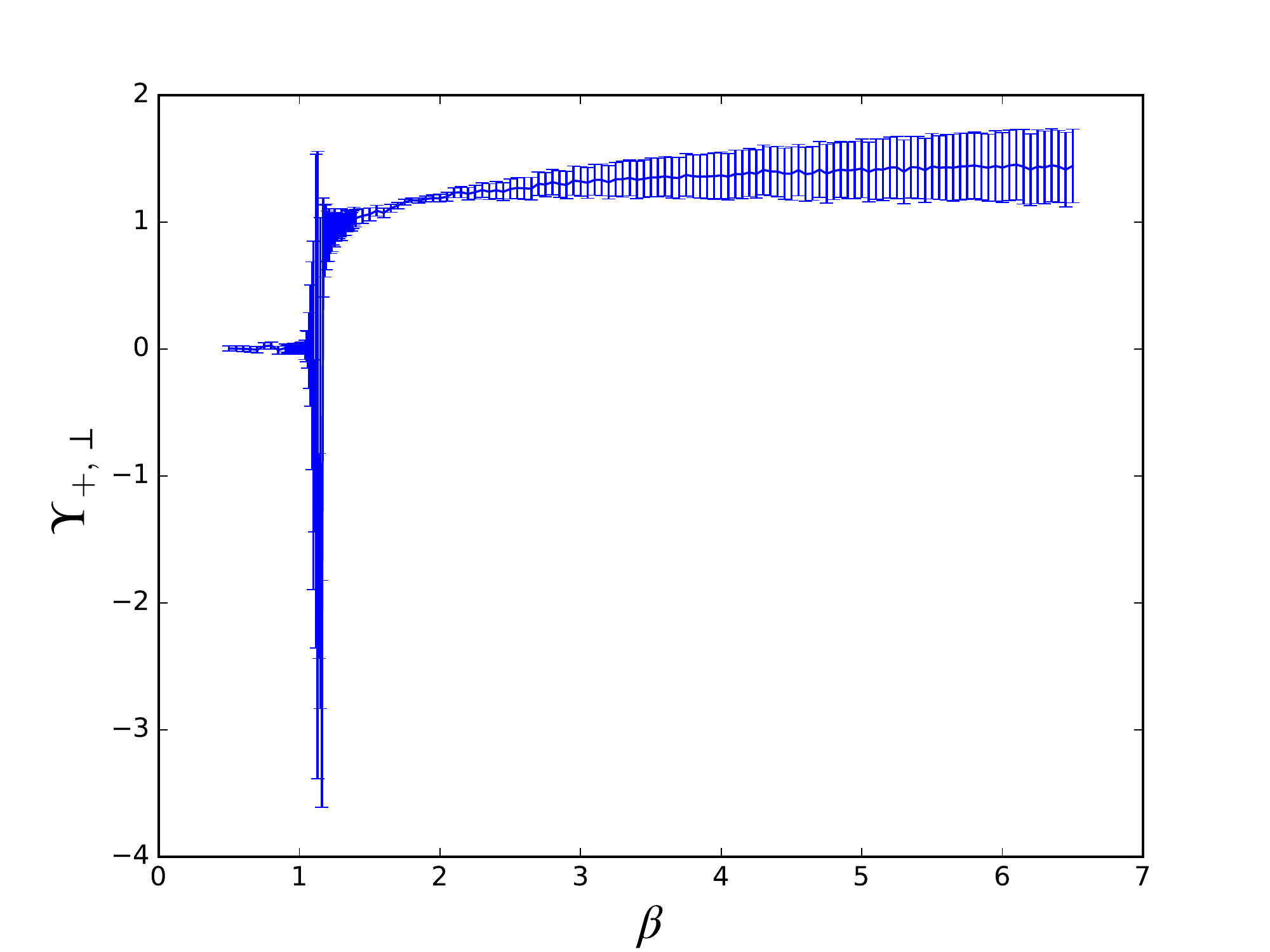}
\label{fig:helinotrapLondon}}
\caption{Phase-sum structure function at the first Bragg peak, $\bv Q$, as a function of $\beta$ for
  system sizes $L\in(16, 24, 32, 40, 48, 56, 64)$ as well as specific heat $C_V/V$ for $L=64$ (top),
  and helicity modulus of the phase-sum variable, $\Upsilon_{+, \perp}$ (bottom), at $\kappa=1.0$ in
  the London limit. The inset of the top panel show the scaling of the peak of the specific heat
  curves for the same system sizes used in the structure function scaling. Note how the drop in the
  exponent $\eta_{\bv Q}$ as well as the jump in the helicity modulus both coincide with the sharp
  peak in the specific heat.}
\label{fig:notrapLondon}
\end{figure}

In order to determine the nature of the thermal phase transition which disorders the modulated
superfluid we measure the helicity modulus of the phase sum variable, the exponent $\eta_{\bv Q}$,
and the specific heat. The helicity modulus is modified compared to the case with no SOC, due to the
extra terms in the Hamiltonian. The value of the exponent $\eta_{\bv q}$ is expected to approach the
limit $1/4$ from below as the critical inverse temperature is approached from
above\cite{Kosterlitz1974} In \cref{fig:notrapnoLondon,fig:notrapLondon} we show the results of the
simulations with and without amplitude fluctuations included, respectively. The top panels show the
specific heat on the left axis, and the value of the exponent $\eta_{\bv Q}$ on the right axis. We
also show the scaling of the specific heat peak in the insets of the top panels, and we find its
exponent to be $0.8(2)$ with amplitude fluctuations included, and $0.66(9)$ in the London limit. In
the bottom panels we show the helicity modulus of the phase sum variable, both of which exhibit a
sharp jump which coincides with the drop in the scaling exponent and the specific heat peak. In both
cases, the sharp peak of the specific heat with its large scaling exponent, the abrupt drop of the
exponent $\eta_{\bv Q}$, and the sharp jump and large error bars of the helicity modulus all point
towards a strong de-pinning transition separating the modulated superfluid phase and the normal
fluid phase. { A KT transition does not fit into the picture presented by
\cref{fig:notrapnoLondon,fig:notrapLondon}, mainly because the specific heat at the KT-transition
temperature has an essential singularity. This singularity is virtually undetectable in numerical
simulations. The fact that we observe such a large and strongly scaling peak in
\cref{fig:cvetanotrapnoLondon,fig:cvetanotrapLondon} rules out a KT-transition almost immediately.
}
The similar behaviours between the two cases of \cref{fig:notrapnoLondon} and
\cref{fig:notrapLondon} suggests that the London model is in fact a good effective model for this
particular transition. We believe the main reason for the pinning is the periodic boundary
conditions applied to the model. This biases the stripes to connect with themselves at the
boundaries of the system, which in turn causes very slow equilibration at the critical point, as
evident in the large error bars of especially the helicity modulus. In particular, fluctuations
associated with shifting or rotating the stripe configurations is particularly hard to resolve in
the Monte Carlo simulations, as these are large scale movements, which in turn are made even more
difficult to resolve with periodic boundary conditions applied.

\begin{figure}
\centering
\includegraphics[width=\columnwidth]{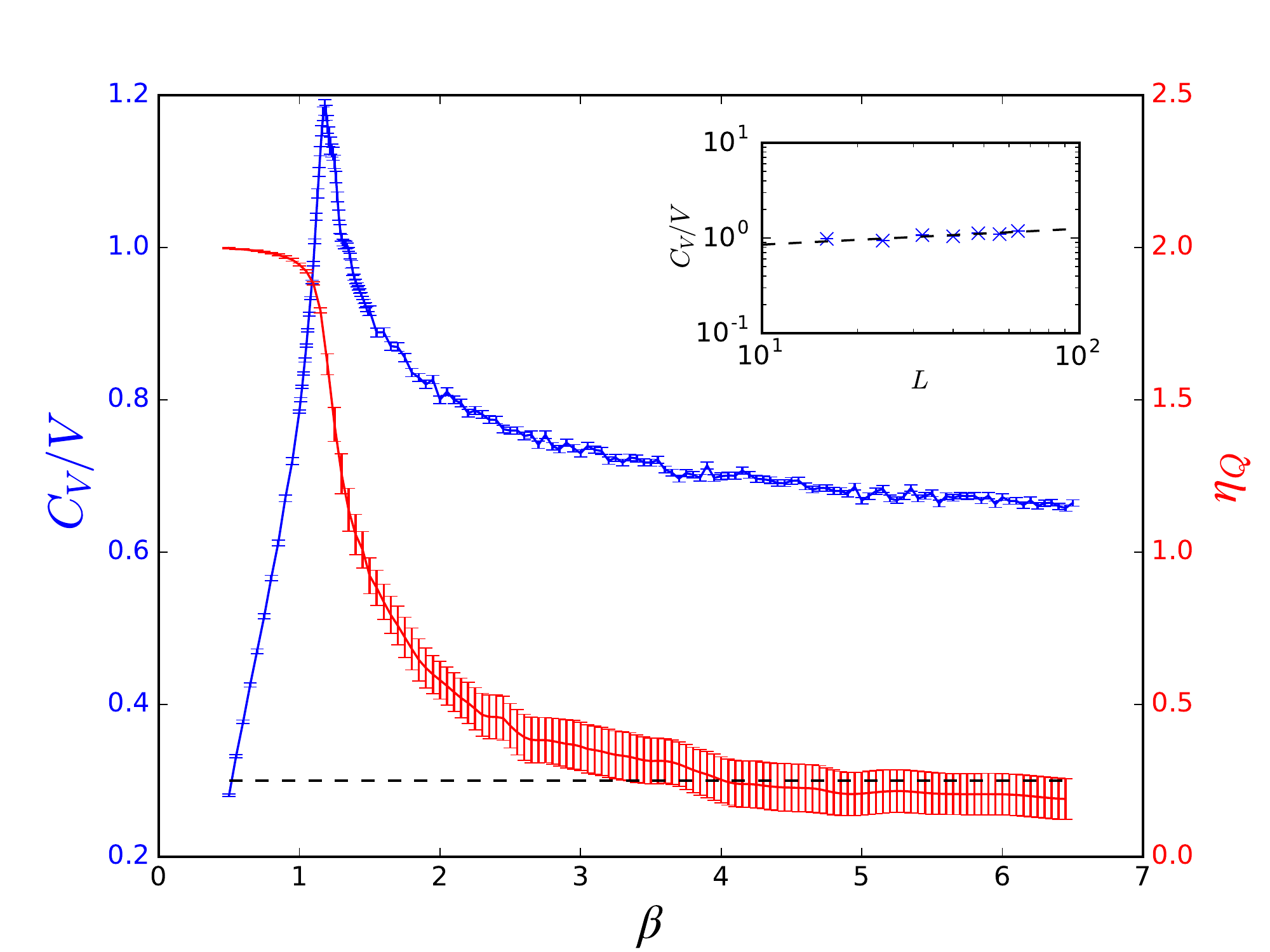}
\caption{Specific heat , $C_V/V$, (left axis) and the exponent $\eta_{\bv Q}$ (right axis). The
  inset shows a finite size scaling of the peak of the specific heat for system sizes $L\in(16, 24,
  32, 40, 48, 56, 64)$ at $\kappa=1.0$ in the London limit with a Thomas-Fermi potential applied.
  The full specific heat curve shown in the main panel is for the largest system sizes simulated,
  $L=64$ Note how the peak of the specific heat curves coincides with the jump in the exponent
  $\eta_{\bv Q}$.}
\label{fig:cvetaLondontrap}
\end{figure}

In an attempt to reduce the pinning effects present in \cref{fig:notrapnoLondon,fig:notrapLondon}
and confirm their origin, we slightly alter the model. Instead of taking the London limit with
$\left|\psi_{\bv r, i}\right|=1\;\forall\;\bv r,i$, we define a Thomas-Fermi trap which decouples
the stripes from the boundaries of the system. Specifically, we fix
\begin{equation}
  \left|\psi_{\bv r, i}\right| =\left\{
    \begin{array}{@{}ll@{}}
      1-\left(\frac{r}{R}\right)^4,&   r<R. \\
      0,&   r>R. \\
    \end{array}\right.
\end{equation}
However, this comes at the cost of not having a well defined helicity modulus.  This is the case for
this particular model, as the decoupling of the stripes from the system boundary is the same as
applying open boundary conditions. The helicity modulus relies on calculating the free energy
difference between the system with periodic boundary conditions, and the system where an
infinitesimal twist is applied to the phases at the boundary\cite{Fisher1973,Li1993}  The simulation
results of the London model in a Thomas-Fermi potential are shown in \cref{fig:cvetaLondontrap}.
Here we show only the scaling of the first order peak in the phase sum structure function and
the specific heat. \cref{fig:cvetaLondontrap} shows that the signs of pinning which we are able to
examine, namely the sharp peak of the specific heat and the sharp drop of $\eta_{\bv Q}$ is greatly
reduced when the Thomas-Fermi potential is present. The specific heat curve still shows a peak which
coincides with the onset of scaling in the structure function, but the height and sharpness of
the peak is reduced. We also find the peak to still exhibit scaling, with an exponent $0.17(4)$, as
shown in the inset of \cref{fig:cvetaLondontrap}. Without the helicity modulus we are unable to
confidently determine the nature of the phase transition, but it is evident that the signs of
pinning is almost removed. In all likelihood, the remaining pinning signatures are associated with
the aforementioned difficulty of moving or rotating entire stripe configurations, and will disappear
in the continuum limit.

\begin{figure}
\centering
\includegraphics[width=\columnwidth]{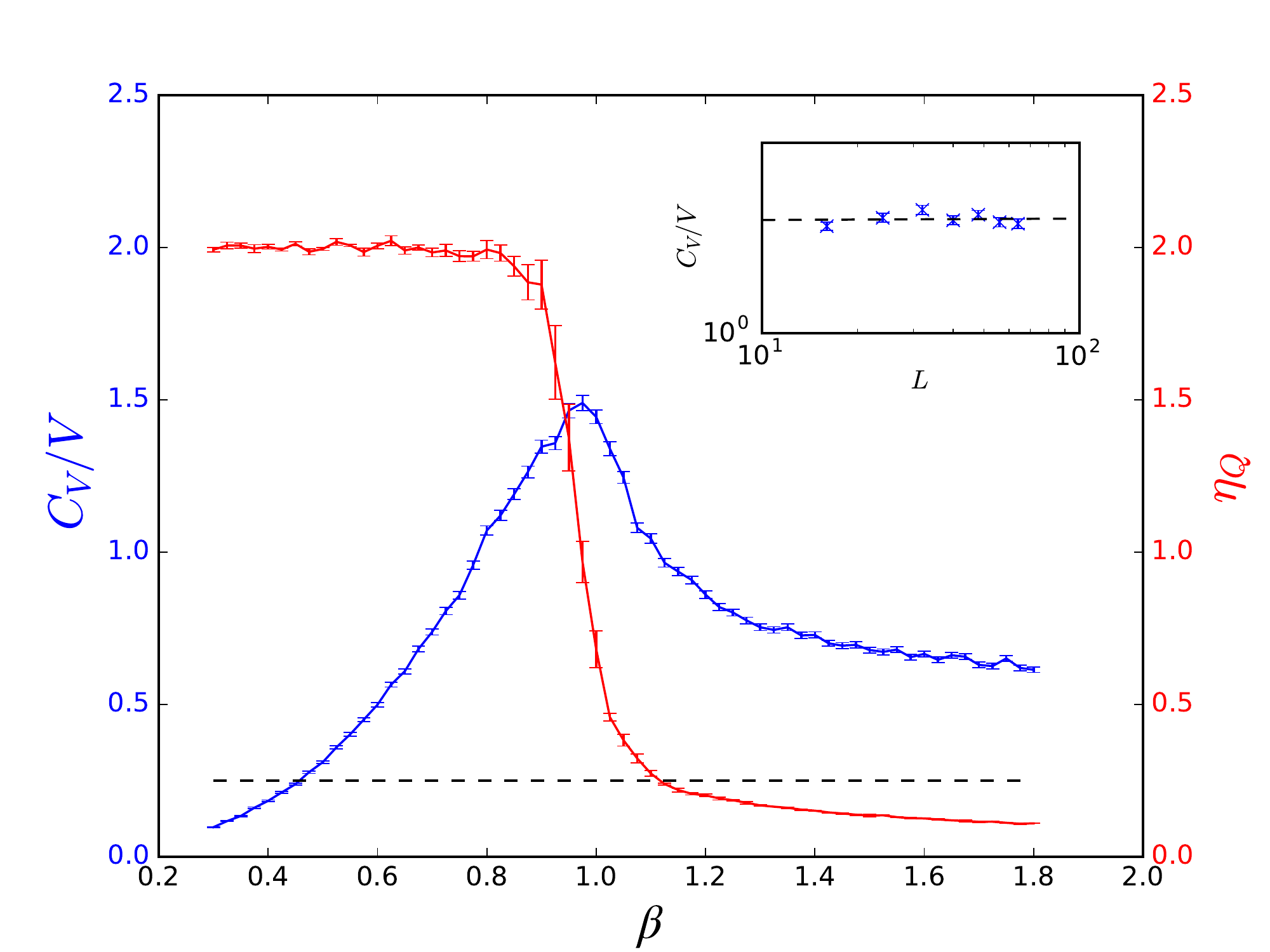}
\caption{Finite size scaling of the height of the $\bv q = 0$ peak in of the structure function
calculated in an XY-model. System sizes $L\in(16, 24, 32, 40, 48, 56, 64)$ have been used. The
exponent grows linearly with temperature to the predicted value of $1/4$ (represented by the dotted
line) at the critical temperature of KT-transition, $\beta_\text{KT}\approx 1.12$.}
\label{fig:cvetaXY}
\end{figure}

As a comparison, we show results for the specific heat and the exponent $\eta_{\bv Q}$ taken from a
simulation of the $2DXY$-model in \cref{fig:cvetaXY}. Here the exponent is measured by performing a
finite size scaling of the height of the $\bv q=0$ peak in the phase structure function. The defining
characteristic which shows that this is a KT-transition is the fact that the exponent $\eta_{\bv Q}$
reaches the limiting value of $1/4$ exactly at the KT-transition temperature,
$\beta_{\text{KT}}\approx 1.12$. We also show the scaling of the specific heat peak, which has an
exponent of $0$ within the errors of our simulation.

Comparing the three different models of
\cref{fig:notrapnoLondon,fig:notrapLondon,fig:cvetaLondontrap}, we may conclude that the
thermal transition from region I of the phase diagram shown in \cref{fig:mfmcnum} into the
disordered phase is a transition from a modulated two-dimensional superfluid phase into a normal
fluid state. The transition has strong de-pinning characteristics when we apply periodic boundary
conditions. These characteristics weakens and we approach a transition consistent with a
KT-transition when we remove the periodic boundary conditions, but we are not able to rigorously
characterize the transition as such due to the lack of a well defined helicity modulus.

\section{Conclusions}
\label{sec:conclusions}

We have studied a model of an imbalanced two-component Bose-Einstein condensate, with and without
spin-orbit coupling in two spatial dimensions, including density-density interactions among the
components. Specifically, we have examined the modulations in the phase-texture of the complex order
parameter components induced by the spin-orbit coupling, its disordering and suppression by thermal
fluctuations and interaction effects, as well as the modulations of the amplitude-texture induced by
a subtle interplay between spin-orbit and inter-component interactions. We also examined the phase
transitions of the model in the parameter regime where SOC is absent.

In the absence of SOC, we found that the phase transition of the model is in the KT universality
class for all values of the inter-component interaction strength we have considered. Here we
observed a KT-transition in the non-suppressed superfluid condensate.  These conclusions are made
based on finite size scaling of the helicity modulus at the transition point, as well as
extrapolation of the negative dip of the fourth order modulus to a non-zero value in the
thermodynamic limit. Both methods strongly indicate a discontinuous jump in the superfluid density
at the critical temperature.

In the presence of SOC, we observed a phase-modulated ground state at finite momenta in Monte-Carlo
simulations. When the inter-component interactions are weaker than the intra-component interactions,
we find that the condensates occupies a single minimum at finite momentum, in agreement with
previous works. This manifests itself as a modulation of the phases of the condensate ordering
fields. For sufficiently strong inter-component interactions and intermediate spin-orbit
interactions, we observed that the spin-orbit induced modulation is completely supressed in favour of
a completely imbalanced condensate. For strong spin-orbit coupling and sufficiently strong
inter-component interactions, however, the total interaction energy is minimized by keeping the
phase-modulation and introducing an additional, staggered modulation of the amplitudes with the same
period. In this phase we observe that the condensate occupies two $q$-vectors of equal magnitude but
opposite alignment.

Finally, we examined the thermal phase transition of the spin-orbit induced plane-wave modulated superfluid ground state
into the normal fluid state in the London approximation. We show that the inclusion of periodic
boundary conditions introduce a strong pinning effect, which weakens as we decouple the stripes from
the edges of the system by applying a Thomas-Fermi potential. In the presence of the potential, we
see signs of a Kosterlitz-Thouless transition, but we are not able to confirm this.

\begin{acknowledgments}
We thank Egor Babaev for useful discussions.  P.~N.~G. was supported by NTNU and the Research
  Council of Norway. A.~S.  was supported by the Research Council of Norway, through Grants
  205591/V20 and 216700/F20, as well as European Science Foundation COST Action MPI1201. This work
  was also supported through the Norwegian consortium for high-performance computing (NOTUR).
\end{acknowledgments}

\appendix

\section{Classification of the KT-transition}
\label{app:KTclass}

The defining characteristic of a Kosterlitz-Thouless transition is the universal jump of
$2/(\pi\beta_c)$ of the superfluid density at the critical temperature, in the thermodynamic limit.
Consider the free energy, where the phase of component $i$ is twisted by an infinitesimal factor
along the $\mu$-direction, $F(\Delta_{i, \mu})$. Technically, this amounts to replacing the phase of
component $i$ by a twisted phase,
\begin{equation}
  \theta_{i, \mathbf{r}}\rightarrow\theta_{i, \mathbf{r}} - r_\mu\Delta_{i, \mu}
\end{equation}
The superfluid density, or helicity modulus, is the second derivative of the free energy
with respect to the twist,
\begin{equation}
  \ev{\Upsilon_{i, \mu}} \equiv \frac{1}{V}\frac{\partial^2 F(\Delta_{i, \mu})}{\partial\Delta_{i, \mu}^2}.
\end{equation}
Similarly, the fourth order modulus is the fourth derivative of the free energy with respect to the
twist,
\begin{equation}
  \ev{\Upsilon_{4, i, \mu}} \equiv \frac{1}{V^2}\frac{\partial^4 F(\Delta_{i, \mu})}{\partial\Delta_{i, \mu}^4}.
\end{equation}
Derivatives of odd order vanish due to symmetry.

In terms of amplitudes and phases of the Ginzburg-Landau theory for a two-component condensate, the
helicity modulus is
\begin{equation}
  V\ev{\Upsilon_{i, \mu}} = \ev{c_{i, \mu}} - \beta\ev{s_{i, \mu}^2},
\end{equation}
while the fourth order modulus is
\begin{align}
  V^2\ev{\Upsilon_{4, i,  \mu}} =  -& 3V^2\beta\ev{(\Upsilon_{i, \mu}-\ev{\Upsilon_{i, \mu}})^2}\nonumber\\
  -& 4V\ev{\Upsilon_{i, \mu}} + 3\ev{c_{i, \mu}} + 2\beta^3\ev{s^4_{i, \mu}},
\end{align}
where we have defined
\begin{align}
  c_{i, \mu} \equiv&\sum_\mathbf{r}\abs{\psi_{i, \mathbf{r}+\boldsymbol\mu}}
  \abs{\psi_{i, \mathbf{r}}}\cos\Big(\theta_{i, \mathbf{r}+\boldsymbol\mu}-\theta_{i,
  \mathbf{r}}\Big),\\
  s_{i, \mu} \equiv&\sum_\mathbf{r}\abs{\psi_{i, \mathbf{r}+\boldsymbol\mu}}
  \abs{\psi_{i, \mathbf{r}}}\sin\Big(\theta_{i, \mathbf{r}+\boldsymbol\mu}-\theta_{i,
  \mathbf{r}}\Big).
\end{align}
This similar to the expressions obtained when considering a 2DXY model. The
amplitude fluctuations only influence the moduli indirectly by weighting the terms in the sums.
Hence, the moduli of each component are coupled indirectly through the potential.

At the critical temperature, the helicity modulus is expected to scale as
\begin{equation}
  \Upsilon_{i, \mu}(L) = \Upsilon(\infty)\Bigg(1 + \frac{1}{2}\frac{1}{\log L + C}\Bigg)
  \label{eq:helifit}
\end{equation}
with system size\cite{Weber1988}. We fit the data at finite size for different values of $\beta$,
and determining at which $\beta$ the best fit is obtained by using the Anderson-Darling test statistic.
This allows an extrapolation of the value of the jump, $\Upsilon(\infty)$, which may be compared
to the KT-prediction. This will also result in an estimate of the critical temperature.

By considering an expansion of the free energy in terms of the phase twist,
\begin{equation}
  F(\Delta_{i, \mu}) - F(0) = \ev{\Upsilon_{i, \mu}}\frac{\Delta^2_{i, \mu}}{2}+\ev{\Upsilon_{4, i,
  \mu}}\frac{\Delta^4_{i, \mu}}{4!}.
\end{equation}
For the system to be stable, the change in the free energy has to greater or equal to zero. If
$\Upsilon_{4, i, \mu}$ is finite and negative in the thermodynamic limit at the critical
temperature, $\Upsilon_{i, \mu}$ cannot go continuously to zero at the critical
temperature\cite{Minnhagen2003}. Therefore, by calculating the negative dip in the fourth order
modulus for increasing system size, a finite value as $L\rightarrow\infty$ signals a discontinuous
jump in the helicity modulus. Furthermore, the temperature at which the dip is located should
converge to the critical temperature. Extrapolation of the location of the dip may therefore be
compared to the above estimate of the critical temperature, as an additional consistency check.
However, this convergence is generally quite slow.

\bibliography{references}

\end{document}